\documentclass[paper]{JHEP}
\usepackage{amsmath}
\usepackage{cite}
\usepackage{epsfig}

\def\epj#1#2#3{{\it Eur. Phys. J. }{\bf C #1} (#2) #3}

\def\cO#1{{\cal{O}}\left(#1\right)}
\def\al{\alpha}
\def\as{\alpha_{\mbox{\scriptsize s}}}

\def\be{\beta}

\def\Real{\mbox{\scriptsize Real}}
\def\out{\mbox{\scriptsize out}}

\def\Ko{K_{\out}}
\def\qq{q\bar{q}}
\def\ee{e^+e^-}
\def\half{\mbox{\small $\frac{1}{2}$}}

\def\cI{{\cal{I}}}

\def\cC{{\cal{C}}}

\def\conf{\delta}
\def\Arg{\mathop{\rm Arg}}
\def\cR{{\cal{R}}}
\def\cF{{\cal{F}}}

\def\abs#1{\left| \: #1 \: \right|}
\def\eps{\epsilon}

\def\om{\omega}
\def\gam{\gamma}

\def\bmu{\bar{\mu}}
\def\bnu{\bar{\nu}}
\def\bKo{\bar{K}_{\out}}
\def\LQCD{\Lambda_{\mbox\scriptsize QCD}}

\def\vk{\vec{k}}
\def\spi{\tau}
\def\vpi{\vec{\spi}}

\def\lrang#1{\left\langle#1\right\rangle}
\def\sgn{\mathop{\rm sgn}}

 \newskip\humongous \humongous=0pt plus 1000pt minus 1000pt
   \newif\ifdtup

\def\la{\mathrel{\mathpalette\fun <}}
\def\ga{\mathrel{\mathpalette\fun >}}
\def\fun#1#2{\lower3.6pt\vbox{\baselineskip0pt\lineskip.9pt
  \ialign{$\mathsurround=0pt#1\hfil##\hfil$\crcr#2\crcr\sim\crcr}}}
\title{QCD analysis of near-to-planar 3-jet events}

\author{
A.~Banfi, G.~Marchesini,\\
Dipartimento di Fisica, Universit{\`a} di Milano--Bicocca
and INFN, Sezione di Milano, Italy} 
\author{
Yu.L.~Dokshitzer,\\
LPT, Universit\'e de Paris XI, Centre d'Orsay, 
France \footnote{on leave from 
PNPI, Gatchina, St.~Petersburg, %188350, 
Russia}}

\author{
G.~Zanderighi.\\
Dipartimento di Fisica, Universit{\`a} di  Pavia
and INFN, Sezione di Pavia, Italy} 

\abstract{Perturbative QCD analysis is presented of the cumulative
  out-of-plane momentum distribution in the near-to-planar $\ee$
  annihilation events, $\Ko\ll Q$. In this kinematical region multiple
  gluon radiation effects become essential. They are resummed with the
  single-logarithmic accuracy, which programme includes the 2-loop
  treatment of the basic radiation and matching with the exact
  $\cO{\as^2}$ result.
  Dedicated experimental analyses of 3-jet event characteristics are
  of special interest for the study of the non-perturbative
  confinement effects.}


\preprint{
     Bicocca--FT--00/02\\
     LPT--Orsay--00--39\\
     FTN/T--2000/06\\
     hep-ph/0004027\\
     April 2000}

\begin{document}

\section{Introduction}
In recent years the standards were established for the perturbative
QCD description of various characteristics of hadron jets
produced in $e^+e^-$ annihilation.
These standards include: 
\begin{itemize}
\item 
  all-order resummation of double- (DL) and single-logarithmic (SL)
  contributions due to soft and collinear gluon radiation effects,
\item 
  two-loop analysis of the basic gluon radiation probability and
\item 
  matching the resummed logarithmic expressions with the exact
  $\cO{\as^2}$ results.
\end{itemize}
Such programmes were carried out for a number of jet shape observables
such as Thrust ($T$) and heavy jet mass ($M_H$)~\cite{thrust},
$C$-parameter~\cite{Cpar} and jet Broadenings (total $B_T$ and
wide-jet broadening $B_W$)~\cite{broad}.

Perturbative description of jets produced in processes other than
$e^+e^-$ annihilation poses more difficulties, as the structure of
jets (both average jet characteristics and distributions) becomes
sensitive to details of the underlying hard interaction. For example,
characteristics of quark jets produced in the current fragmentation
region in Deep Inelastic Scattering (DIS) depend on the Bjorken $x$.
In spite of these complications a steady progress is being made in
this domain as well~\cite{thrustDIS}.

Given a (relative) perfection of perturbative QCD technologies, it
became possible to aim at {\em deviations}\/ of the measured hadronic
characteristics from the corresponding perturbative predictions, in a
search for genuine non-perturbative confinement effects.  As is well
known, these deviations, for a broad variety of jet observables,
amount to sizable $1/Q$ corrections, with $Q$ the hardness scale (the
total $e^+e^-$ annihilation energy), for reviews see~\cite{1/Q,Blois}.

As far as physics of $e^+e^-$ annihilation is concerned, till now
these developments, both in the perturbative and non-perturbative
sectors, were confined to two-jet ensembles which constitute the bulk
of events.  Little (if any) attention has been paid to multi-jet
ensembles, in particular to three-jet events.  In spite of being
obviously more rare ($\sigma_{N}/\sigma_{tot}\propto \as^{N-2}$, with
$N$ the number of jets), selected multi-jet configurations are of
special interest as they, so to speak, are subject to {\em more}\/
quantum mechanics than unrestricted (mainly two-jet) hadron production
events.

Indeed, in the latter case it is known that the gross inclusive
features of particle production can be described in {\em probabilistic
  terms}\/ by imposing {\em angular ordering}\/~\cite{AO} on
successive $1\to2$ intra-jet parton decays.  It suffices to implement
{\em strict}\/ angular ordering, $\theta_{i+1}\le\theta_i$, the proper
running coupling, $\as(k_\perp)$, and the standard parton splitting
probabilities
to ensure the {\em next-to-leading}\/ accuracy of the perturbative
description of inclusive energy spectra, mean multiplicities and the
multiplicity distribution~\cite{SAO}.

Thus, a transparent and powerful probabilistic technology exists for
predicting (with next-to-leading accuracy) {\em inclusive}\/
observables.  At the same time, the very selection of events (say,
three-jet events) necessarily makes a measurement ``less inclusive''
and destroys, generally speaking, the probabilistic picture, and one's
hold on the SL accuracy with it.

Of special interest for two-jet configurations is the kinematical
region of two {\em narrow}\/ jets ($1-T,M_H^2/Q^2,C,B_{T,W}\ll1$)
where multiple radiation (Sudakov suppression) effects are essential.
An analogue of this region for three-jet events is near-to-planar
kinematics.  Three hard partons --- $\qq$ and a gluon $g$ --- form a
plane.  Secondary gluon radiation off the three-prong QCD antenna
brings in aplanarity.  In what follows we choose the out-of-plane
transverse momentum $\Ko$ as an aplanarity measure.
The distribution of events in $\Ko$ is subject to DL suppression in
the region of small aplanarity, $\Ko/Q\ll1$, where normal accompanying
radiation is vetoed.

The corresponding suppression factor, and thus the resulting
$\Ko$-distribution is easy to predict in a DL approximation which
takes care of the leading contributions $\cO{[\as\ln^2(Q/\Ko)]^n}$ in
all orders while disregarding subleading SL corrections of the order
of $[\as\ln(Q/\Ko)]^m$.  The answer is given simply by the product of
three proper QCD Sudakov factors, $F^2_q\cdot F_g$, which veto
radiation of gluons with out-of-plane momentum components exceeding
$\Ko$, off the three hard partons treated as independent emitters.

The DL approximation is known to be too rough to be practically
reliable.  Improving it proves to be nontrivial a quest: at the level
of SL terms the {\em geometry}\/ of the underlying three-jet
configuration enters the game: in addition to {\em intra-jet}\/ parton
multiplication one has to take into consideration {\em inter-jet}\/
particle production.  The latter, however, does not admit
``classical'' probabilistic interpretation: gluon radiation in-between
jets results from the coherent action of all three elements of the
antenna.

This is a general feature which complicates the analyses of other, more
simple, characteristics of three-jet ensembles as well.  For example,
contribution of coherent inter-jet particle flows enters, at the level
of the next-to-leading $\cO{\sqrt{\as}}$ correction, into perturbative
prediction of the mean particle multiplicity in three-jet events,
making it event-geometry-dependent~\cite{mm3jet}.

In this paper we attempt, for the first time, the all-order
perturbative analysis of the $\Ko$-distribution in three-jet $e^+e^-$
annihilation events, aiming at SL accuracy.
In what follows we will single out and resum logarithmically-enhanced
DL and SL contributions and systematically neglect relative
corrections of the order $\cO{\as}$.  The latter belong to the
non-logarithmic normalization factor (``coefficient function'')
$1+c{\as}+\ldots$, whose first coefficient, $c$, can be found by
comparing the approximate resummed result with numerical calculation
based on the exact $\as^2$ matrix element~\cite{BSZ}.

To accommodate all essential SL contributions one has
\begin{enumerate}
\item 
  to take into account soft inter-jet gluon radiation,
\item
  to analyse corrections due to hard intra-jet parton decays,
\item
  to define an event plane for a multi-parton system and properly
  treat kinematical effects due to parton recoil,
\item
  to prove, for three-jet environment, soft gluon exponentiation (at
  the two-loop level) and the prescription for the argument of the
  running coupling that enters the basic gluon emission probability.
\end{enumerate}
In the present paper we address these issues.

The answer we derive has the following key features:
\begin{itemize}
\item
  Kinematical constraints which determine an event plane are rather
  complicated but can be resolved with a help of multiple
  Fourier--Mellin representation which allows for exponentiation of
  multiple radiation in the parameter space.
\item 
  Multiple soft radiation off the three-parton system can be resummed
  and exponentiated in terms of three colour dipoles that together
  determine the colour structure of (and accompanying particle flows
  in) the event.
\item 
  For the cumulative $\Ko$-distribution (as well as for other
  sufficiently inclusive observables) inclusive treatment of the
  two-parton decay of a gluon emitted by the $\qq\,g$ system results
  in the running of the coupling constant.
\item
  The running $\as$ which describes the intensity of gluon emission
  off each dipole ($qg$, $\bar{q}g$, $\qq$) depends on the invariant
  transverse momentum of the gluon with respect to two partons that
  form the corresponding dipole.
\item 
  The exponent can be represented as a sum of three basic parton
  ``radiators'' each of which describes one-gluon emission off a
  single hard parton, weighted by the colour factor of this parton.
  This radiation can be treated as independent provided a proper
  hardness scale is ascribed to the parton radiator.
\item 
  Essential SL contribution (``hard'' intra-jet and coherent inter-jet
  corrections) can be conveniently embodied into the scales.  Effects
  of the large-angle soft (inter-jet) radiation make these scales
  event-geometry-dependent and different for each of the three primary
  partons.
\item 
  The structure of the hardness scales of the parton radiators
  entering in the total $\Ko$-distribution has a clear geometrical
  interpretation: the scale $Q_a$ for the parton $a$ is proportional
  to the invariant transverse momentum of this parton, $p_{ta}$, with
  respect to the hyper-plane formed by the other two hard partons.
\item
  Equivalent expressions for the radiators of the total
  $\Ko$-distributions can be constructed which smoothly interpolate
  between 3- and 2-jet configurations.
\end{itemize}
As has been already said, kinematics of three-jet observables
complicates the analysis. As a result, the final expressions are
rather cumbersome as they involve 4- and 5-dimensional integrals.
 
The kinematical constraints, and thus the final formulae, are somewhat
simpler for the distribution of $\Ko$ accumulated in the {\em right}\/
hemisphere, i.e. the one that has the smallest transverse momentum
with respect to the thrust axis (for a review see ~\cite{l3col}).

Quantitative analysis of the predictions and numerical results are
discussed in a separate publication~\cite{BSZ}.

The paper is organised as follows.\\
In Section~\ref{sec:aco} we define the aplanarity measure $\Ko$,
derive kinematical relations defining the event plane and discuss
resummation of soft gluon radiation for near-to-planar
3-jet events.\\
Section~\ref{sec:rig} is devoted to perturbative analysis of the
$\Ko$-distribution in the right hemisphere.  Here we develop technique
for analysing and embodying all necessary SL
corrections into the all-order perturbative result.\\
In Section~\ref{sec:tot} we apply this technique to derive, with SL
accuracy, the perturbative prediction for the total $\Ko$-distribution
in 3-jet events with given kinematics (thrust $T$ and thrust-major
$T_M$).

Technical details are confined to Appendices.

\section{Aplanarity and soft parton resummation\label{sec:aco}}

We consider $\ee$ annihilation events with 
almost planar configuration of final state particles. 
Such event are characterised by the inequality
\begin{equation}
  \label{eq:3jetkin}
  T\sim T_{M} \gg T_{m}\>, 
\end{equation}
with $T$ the thrust, $T_M$ and $T_m$ the so-called thrust-major and
thrust-minor. We study the distribution of $T_m$. We shall refer
to $\ee$ events in this phase space region as $3$-jet events.

Thrust is defined as
\begin{equation}
  \label{eq:thrust}
  T\,Q = \max_{\vec{n}} \left\{ \sum_{h} \abs{\vec{n}\vec{p}_h}
    \right\} \>=\>  \sum_{h} \abs{\vec{n}_T\vec{p}_h} \>=\>
\sum_{h} \abs{{p}_{hz}} \>,
\end{equation}
where the sum runs over all particles $h$ produced in a given event
with the total center of mass energy $Q$. 
Hereafter we choose the $z$-axis to lie along the thrust
axis, $\vec{n}_T$.
The thrust-major is defined analogously; it maximises the sum of the
particle momenta projections in the two-dimensional plane orthogonal
to the thrust axis:
\begin{equation}
  \label{eq:thrustM}
  T_M\,Q = \max_{\vec{n}\vec{n}_T=0} 
   \left\{ \sum_{h} \abs{\vec{n}\vec{p}_h}
    \right\} \>=\>  \sum_{h} \abs{\vec{n}_M\vec{p}_h} 
  \>=\> \sum_{h} \abs{{p}_{hy}} \>.
\end{equation}
The direction $\vec{n}_M$ we shall identify with the $y$-axis.
Finally, for the thrust-minor we have
\begin{equation}
  \label{eq:thrustm}
  T_m\,Q = \sum_{h} \abs{{p}_{hx}} \>\equiv\> \Ko \>.
\end{equation}
We choose $\vec{n}_T$ in such a way that the most energetic particle 
in the event has a positive $z$-component and $\vec{n}_M$ in such 
a way that the second most energetic particle has a positive $y$-component. 
Hereafter we attribute $p_h$ to the {\bf right}
hemisphere $C_R$ ({\bf left} hemisphere $C_L$) if $p_{hz}>0$
($p_{hz}<0$). Similarly $p_h$ is in the {\bf up} hemisphere $C_U$
({\bf down} hemisphere $C_D$) if $p_{hy}>0$ ($p_{hy}<0$).
We shall also consider separately the right-hemisphere 
cumulative out-of-plane transverse momentum: 
\begin{equation}
  \label{eq:KoutRL}
    \Ko^{R} \>\equiv\> \sum_{h\in R} \abs{{p}_{hx}}\>.
\end{equation}
It can be easily shown that from the definition
(\ref{eq:thrust}--\ref{eq:thrustM}) the kinematical constraints
follow:
\begin{equation}
\label{eq:RLUD}
\begin{split}
  \sum_{h\in R} \vec{p}_{ht} = \sum_{h\in L} \vec{p}_{ht} =0 \>, 
\qquad
  \sum_{h\in U} p_{hx} = \sum_{h\in D} p_{hx} =0\>,
\end{split}
\end{equation}
where $\vec{p}_{t}$ is the two-dimensional vector transversal to the $z$-axis.
We introduce the three-dimensional vector $\vec{P}_1$ and the
two-dimensional vector $\vec{P}_{2t}$ that define the ``event plane''
$\{z,y\}$:
\begin{equation}
  \label{eq:P0}
\vec{P}_1 = (P_{1x},P_{1y},P_{1z})=(0,0,TE)\>, \quad 
\vec{P}_{2t}=(P_{2x},P_{2y})= (0,T_M E)\>, \quad Q=2E\>.
\end{equation}
From the definition of $T$ and $T_M$ we have
\begin{equation}
  \label{eq:P12}
\vec{P}_1\>=\>\sum_{h\in R}\vec{p}_{h}\>,\qquad
\vec{P}_{2t}\>=\>\sum_{h\in U}\vec{p}_{ht}\>.
\end{equation}
In what follows we study the distribution of events in the cumulative
out-of-plane transverse momentum $\Ko$ defined in \eqref{eq:thrustm}.
The integrated $\Ko$-distribution is defined as
\begin{equation}
\label{eq:Sig}
\frac{d\sigma(\Ko)}{dTdT_{M}}= Q^5 \sum_m\int {d\sigma_m}\>
\Theta\left(\Ko\!-\!\sum_{h=1}^{m} |p_{hx}|\right)
 \delta^{3}\left(\sum_{h\in R}\vec{p}_{h}\!-\! \vec{P}_1\right)
 \delta^{2}\left(\sum_{h\in U}\vec{p}_{ht}\!-\!\vec{P}_{2t}\right),
\end{equation}
where $m$ denotes the number of final particles in an event.  The last
two delta-functions fix the event plane and the theta-function defines
the observable.  Analogously we define the {\em right}\/ distribution
by restricting the sum over particles in the theta-function of
\eqref{eq:Sig} to those belonging to the right hemisphere (see
\eqref{eq:KoutRL}).

\subsection{Jet momenta and hard parton recoil}
At the parton level the events in the region \eqref{eq:3jetkin} can be
treated as three-jet events generated by a system of energetic quark,
antiquark and a gluon accompanied by an ensemble of soft partons.

At the Born level, $\cO{\as}$, in the absence of accompanying
radiation the 3-parton system is truly planar, $T_m\equiv0$. The
kinematical configuration of $q, \bar{q}$ and $g$ treated as massless
partons is then uniquely fixed by the values of $T$ and $T_M$.
Denoting by $P_1,P_2$ and $P_3$ the energy ordered parton momenta,
$P_{10}>P_{20}>P_{30}$, we have $\vec{P}_1$ lying along the thrust
axis. The vectors $\vec{P}_1$ and $\vec{P}_{2t}=-\vec{P}_{3t}$ are
given by the event plane momenta defined in \eqref{eq:P0}.

There are various kinematical configurations of the Born system.  The
three parton momenta $P_q, P_{\bar q}, P_g$ can belong to three
configurations $\cC_{\conf}$, namely (see Figure~\ref{fig:conf})
\begin{equation}
  \label{eq:conf}
\begin{split}
 (P_1,P_2,P_3)\>&=\> (P_g,P_q, P_{\bar q}) \>\Rightarrow\>   \cC_1 
\\&= \>(P_q,P_g, P_{\bar q}) \>\Rightarrow\>   \cC_2
\\&= \>(P_q,P_{\bar q},P_g) \>\Rightarrow\>   \cC_3\>.
\end{split}
\end{equation}
Notice that the index $\conf$ labelling the configuration 
$\cC_{\conf}$ coincides with the index of the gluon momentum.
(Interchanging the quark and the antiquark does not affect the
accompanying radiation and the distributions under study.) 
\EPSFIGURE[ht]{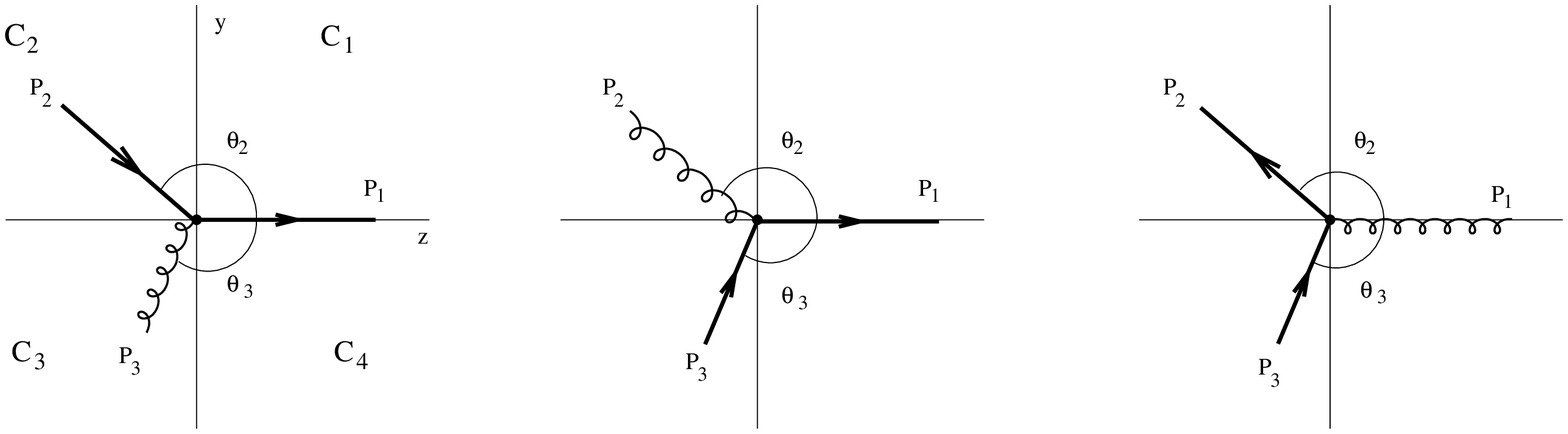,width=1.0\textwidth} {The three Born
      configurations $\cC_3,\cC_2$ and $\cC_1$ for $T=0.75$ and
      $T_M=0.48$ ordered according to decreasing probability.  The
      thrust $T$ and thrust major $T_M$ are along the z- and y-axis
      respectively. The four regions $C_{\ell}$ in the phase space are
      indicated.
\label{fig:conf}}
For three massless partons the values of $T$ and $T_M$ are restricted
to the region
\begin{equation}
  \label{eq:kin3}
  \frac{2\>(1-T)}{T}\sqrt{2T-1}\> < \>T_M < \sqrt{1-T}\>.
\end{equation}
(For the kinematics of the Born system momenta see
Appendix~\ref{App:kin}.)  In the following we will analyse the
distribution inside this region --- the 3-jet region --- in which
three skeleton parton momenta $P_a$ can be reconstructed 
from the $T$ and $T_M$ values.

Beyond the Born approximation, in the presence of secondary gluon
radiation, $T_m$ is no longer vanishing. The $x$-components of
bremsstrahlung gluon momenta are logarithmically distributed over a
broad range, so that $\lrang{T_m}\sim \as T_M\sim \as T$. 

In the region \eqref{eq:3jetkin} the PT expansion develops
logarithmically enhanced contributions which need to be resummed in
all orders.  Our aim is to perform this resummation with SL
accuracy. This means that we shall keep both DL, $\as\log^2 T_m$, and
SL, $\as\log T_m$, contributions to the exponent (``radiator'', see
below) while neglecting non-logarithmic corrections $\cO{\as}$.  With
account of the running coupling effect, the DL and SL contributions
formally expand into series of terms $\as^n\log^{n+1}T_m$ and
$\as^n\log^n T_m$, respectively.

It is important to stress that exponentiation of the SL correction
$\as\log T_m$ makes sense only if it is supported by the calculation 
of the non-logarithmic ``coefficient function''.
Indeed, a ``cross-talk'' between an $\cO{\as}$ correction to the
overall normalization of the resummed distribution and the leading DL
term, symbolically,
\begin{equation}
  \label{eq:expans}
(1+c\cdot\as)\!\times\! \exp\left\{ \as\log^2 + s\cdot \as\log\right\}
= 1+\!\ldots\! + c\cdot\as\times \as\log^2 + \frac{s^2}{2!}(\as\log)^2 +\! \ldots,  
\end{equation}
gives rise to the correction of the same order as the SL term squared.
The coefficient $c$ is not known analytically. It can be determined
by comparing the $\cO{\as^2}$ term of the log-resummed expression with
the result of a numerical calculation based on the exact $\as^2$ matrix
element~\cite{BSZ}. 

We denote by $p_a$ the momenta of the three hard partons,
$q$, $\bar{q}$, $g$,
that in general no longer lie in the event plane defined by the vectors
\eqref{eq:P0}. In the 
region \eqref{eq:3jetkin} they differ from $P_a$ by ``soft recoil
parts'' $q_a$,
\begin{equation}
  \label{eq:Pa}
p_a=P_a+q_a\>,  
\end{equation}
whose $x$-components contribute to $\Ko$ together with the soft
parton momenta $k_i$:
\begin{equation}
  \label{eq:KT}
  \Ko=|q_{1x}|+|q_{2x}|+|q_{3x}|+\sum|k_{ix}|\>.
\end{equation}
To SL accuracy, as in the case of the broadening distribution
\cite{broad}, it suffices to keep the recoil momenta $q_a$ only in the
phase space, in particular their contribution to $\Ko$ and to the
constraints defining the event plane (see \eqref{eq:Sig}).  At the
same time, $q_a$ can be neglected in the radiation matrix element,
that is, in the emission distributions we can substitute the skeleton
momenta $P_a$ for actual parton momenta $p_a$.  We remark that, as in
the case of jet broadening~\cite{broadNP}, for the study of {\em
  non-perturbative}\/ power-suppressed corrections the approximation
$p_a\to P_a$ in the radiation matrix element is not valid~\cite{KoNP}.

In order to resum the PT series for \eqref{eq:Sig} in the 
region \eqref{eq:3jetkin} we need to use the factorization property of
the soft radiation matrix element and to factorize the multi-parton
phase space. We discuss these points in succession after giving a
brief description of the kinematics. For more details see
Appendix~\ref{App:kin}.

\subsection{Kinematics of soft or collinear emission}
Introducing the recoil momenta $q_a$ defined in \eqref{eq:Pa},
the right, left and total $\Ko$ are given by
\begin{equation}
  \label{eq:Ko}
  \Ko^R=|q_{1x}|+\sum_R|k_{ix}|\>,\quad
  \Ko^L=|q_{2x}|+|q_{3x}|+\sum_L|k_{ix}|\>,\quad
  \Ko=\Ko^R+\Ko^L\>.
\end{equation}
The event plane momenta \eqref{eq:P0} are given by
\begin{equation}
  \label{eq:kin}
  \begin{split}
\vec{P}_1\>=\>\vec{p}_1+\sum_R\vec{k}_i\>,\qquad 
\vec{P}_{2t}\>=\>\vec{p}_{2t}+\vec{p}_{1t}\vartheta(p_{1y})
+\sum_{U}\!\vec{k}_{it}\,.
\end{split}
\end{equation}
In the region in which the emitted partons are soft or collinear 
the components of $q_a$ that accompany {\em
  non-vanishing}\/ components of $P_a$ can be neglected in the
calculation.  Therefore we consider only those four components that
vanish in the Born approximation namely, the $y$-component of the leading
parton momentum, $q_{1y}=p_{1y}$, and the $x$-components
$q_{ax}=p_{ax}$, $a=1,2,3$. 
Then, from the kinematical relations \eqref{eq:kin} and
\eqref{eq:KoutRL} we have
\begin{equation}
  \label{eq:softq}
  \begin{split}
  &q_{1y}+  \sum_R k_{iy}=
   q_{1x}+  \sum_R k_{1x}=0\>,\\
  &q_{2x}+q^+_{1x}+  \sum_U k_{1x}=   q_{3x}+q^-_{1x}+  \sum_D k_{1x}=0\>,
\qquad {q}^{\pm}_{1x}={q}_{1x}\vartheta(\pm q_{1y})\>.
  \end{split}
\end{equation}
Notice that this implies the following kinematical relations
\begin{equation}
\begin{split}  
  \big\{q^+_{1x}+\sum_{i\in C_1}k_{ix}\big\}
=-\big\{q^-_{1x}+\sum_{i\in C_4}k_{ix}\big\}
=-\big\{{q}_{2x}+\sum_{i\in C_2}{k}_{ix}\big\}
= \big\{{q}_{3x}+\sum_{i\in C_3}{k}_{ix}\big\}.
\end{split}
\end{equation}

\subsection{Matrix element factorization}
At the parton level the distribution is given by the sum
of the partial cross sections 
\begin{equation}
\label{eq:dsigma}
  d\sigma_n=\>\frac1{n!}\>M_{n}^2 \cdot d\Phi_n\>,
\end{equation}
where $M_{n}$ is the matrix element for the emission of $n$ secondary
partons off the $\qq$, $g$ system, and $d\Phi_n$ the corresponding
phase-space factor,
\begin{equation}
\begin{split}
\label{eq:dPn}
d\Phi_n\>=\>
(2\pi)^4\delta^4\!\!\left(\!\sum_a p_a\!+\!\sum_i k_i\!-\!Q\!\right)
\prod_{i=1}^n [dk_i] \cdot  
 \prod_{a=1}^3 \frac{d^3p_a}{(2\pi)^3 2E_a}\>, 
\qquad [dk]= \frac{d^3k}{\pi \om}\>.
\end{split}
\end{equation}
As it is explained in detail in Appendix~\ref{App:d1a} the contribution of 
collinear non soft emission is process independent and can be 
taken into account {\it a posteriori} introducing in the computed 
distribution the hard part of the splitting functions. 

Therefore the accompanying partons $k_i$ are assumed to be soft and
their distribution can be treated as independent.  Assembling
together the contributions from the three configurations $\cC_{\conf}$
defined in \eqref{eq:conf}, the distribution can be presented as
\begin{equation}
\label{eq:Mn2}
  M_{n}^2 \>=\> \sum_{\conf=1}^3\>M_{0}^2(\cC_{\conf}) \cdot 
\prod_{i}^n W_{\conf}(k_i)\>,
\end{equation}
where $M_{0}(\cC_{\conf})$ is the Born $\qq g$ matrix element and $W_{\conf}$ 
is the distribution of the soft gluon radiation off the hard three-parton
antenna in the momentum configuration $\cC_{\conf}$.  
Here we discuss the real emission contribution to $M_{n}$; the virtual
corrections will be accommodated later.

For the configuration  $\conf=3$, for example, 
the squared Born matrix element reads
\begin{equation}
  \label{eq:Born}
M_{0}^2(\cC_3) \>=\> \frac{C_F\as}{2\pi} \frac{x_1^2+x_2^2}{(1-x_1)(1-x_2)}\>,
\qquad x_a\equiv \frac{2P_aQ}{Q^2}\>,
\end{equation}
where $P_3$ is the gluon momentum and $P_1, P_2$ the quark-antiquark
momenta (see \eqref{eq:conf}).  For this configuration the single soft
gluon radiation pattern is given, at the one-loop level, by
\begin{equation}
  \label{eq:W3}
W_{3}(k)= C_F \,w_{12} +\frac{N_c}{2}\big( w_{13}+w_{23}-w_{12}\big)
= \frac{N_c}{2}\left( w_{13}+w_{23}-\frac{1}{N_c^2}w_{12}\right).  
\end{equation}
Here $w_{ab}$ is the standard two-parton antenna of the $ab$-dipole,
which, within the normalization convention prescribed by
\eqref{eq:dPn}, is given by
\begin{equation}
  \begin{split}
\label{eq:wij}
w_{ab}(k) =  \frac{\as}{\pi}\frac{(P_aP_b)}{2(P_ak)(kP_b)}
= \frac{\as}{\pi k_{t,ab}^2}\>.
\end{split}
\end{equation}
Here $k_{t,ab}$ is the invariant gluon transverse momentum with
respect to the hyper-plane defined by the $P_a,P_b$ momenta.

The first term $C_F w_{12}$ in \eqref{eq:W3} is the ``Abelian''
contribution describing soft gluon emission off the $\qq$ pair.  The
second term proportional to $N_c$ is its ``non-Abelian'' counterpart
that describes radiation off the hard gluon $P_3$.  Similar
expressions for two other kinematical configurations ($\conf=1,2$) is
straightforward to write down by properly adjusting the parton
indices,
\begin{equation}
  \label{eq:altre}
  \begin{split}
&W_{1}(k)= \frac{N_c}{2}\left(
  w_{13}+w_{12}-\frac{1}{N_c^2}w_{23}\right),\\
&W_{2}(k)= \frac{N_c}{2}\left(
  w_{23}+w_{12}-\frac{1}{N_c^2}w_{13}\right).
\end{split}
\end{equation}
To reach SL accuracy it is necessary to treat multi-parton emission at
the two-loop level.  This involves allowing secondary gluon to split
into two gluons or into a $\qq$ pair.  In principle, perturbative
analysis of a system consisting of three hard partons and two softer
partons (with comparable energies) can be found in the literature
(see, e.g.,~\cite{Catani}).  However, to the best of our knowledge,
the most important feature of the result has never been explicitly
stressed, namely, that the colour structure and geometrical properties
of emission of a soft two-parton system off a 3-jet ensemble is {\em
  identical}\/ to those for single gluon radiation.

It can be shown~\cite{BDMZ} that after subtracting the uncorrelated
radiation of two soft gluons, $W_3(k_1)W_3(k_2)$, the correlated
two-parton production is given by the expression
\begin{equation}
  \label{eq:correl}
  W_3^{(2)}(k_1,k_2) = 
  C_Fw_{12}^{(2)}(k_1,k_2) 
  + \frac{N_c}{2}\left(
  w_{13}^{(2)}(k_1,k_2)+w_{23}^{(2)}(k_1,k_2) -w_{12}^{(2)}(k_1,k_2)\right),
\end{equation}
where $w_{ab}^{(2)}$ is the standard distribution known from
the two-loop analysis of 2-jet event shapes
(the first term on the right-hand side of \eqref{eq:correl}, 
see~\cite{DLMS}). 
It describes decay into $\qq$- or $gg$-pair of a virtual parent gluon 
radiated, in our case, by one of the three two-parton dipoles $ab$.
Given this remarkably simple dipole structure, the analysis of the
two-loop effects in 3-jet events reduces to the known 2-jet case.  
 
As shown in Ref.~\cite{DLMS},
for a sufficiently inclusive observable (and the $\Ko$-distribution under
interest in particular) 
the two-loop refinement results in (and reduces to) substituting the
proper argument for the running coupling 
describing the intensity of the parent gluon emission,
\begin{equation}
  \label{eq:W3'}
   w_{ab}(k)=\frac{\as}{\pi k_{t,ab}^2}\>\Longrightarrow\>
   w_{ab}(k)+\int dk_1dk_2\delta(k-k_1-k_2)\,w_{ab}^{(2)}(k_1,k_2) 
\simeq \frac{\as{(k_{t,ab})}}{\pi k^2_{t,ab}}\>.
\end{equation}
Here $k_{t,ab}$ is the invariant transverse momentum defined in \eqref{eq:wij}
and $\as$ is taken in the physical scheme~\cite{CMW}.  
It is worthwhile to notice
that the arguments of the coupling are different for the three dipoles
that participate in gluon radiation according to~\eqref{eq:W3}.

We remark that at the level of the leading power-suppressed {\em
  non-perturbative}\/ correction the r{\^o}le of two-loop effects is more
involved as they give rise to the Milan factor~\cite{DLMS,Milan,Blois}.

\subsection{Phase space factorisation}
We discuss here factorization of the phase space in the soft region,
which is needed for resummation.
The phase space factor reads
\begin{equation}
\begin{split}
\label{eq:ps}
d\Gamma_n 
&= d\Phi_n\cdot 
\delta^3\!\!\left(\!\vec{p}_1 \!+\! \sum_R\vec{k}_i \!-\!\vec{P}_1\right)
\delta^2\!\!\left(\!\vec{p}_{2t} \!+\! 
\vec{p}_{1t}\vartheta(p_{1y}) \!+\! \sum_{U}\!\vec{k}_{it}
\!-\!\vec{P}_{2t}\!\right) \\
&=\prod_{i=1}^n [dk_i] \cdot  
 \prod_{a=1}^3 \frac{d^3p_a}{(2\pi)^3 2E_a} \cdot D_n\>, 
\qquad [dk]= \frac{d^3k}{\pi \om}\>,
\end{split}
\end{equation}
where the factor $D_n$ takes care of the kinematical relations,
\begin{equation}
\label{eq:Dn}
D_n\!=\! (2\pi)^4
\delta^4\!\!\left(\!\sum_a p_a \!+\! \sum_i k_i \!-\!Q\!\right)
\!\delta^3\!\!\left(\!\vec{p}_1 \!+\! \sum_R\vec{k}_i \!-\!\vec{P}_1\right)
\!\delta^2\!\!\left(\!\vec{p}_{2t} \!+\! 
\vec{p}_{1t}\vartheta(p_{1y}) \!+\! \sum_{U}\!\vec{k}_{it}
\!-\!\vec{P}_{2t}\!\right).
\end{equation}
The first delta-function stands for the energy--momentum
conservation, while the last two (three- and two-dimensional) 
delta-functions set the event plane.

Now we single out from $d\Gamma_n$ small recoil components of hard
partons momenta, $q_{1y}=p_{1y}$ and the three $q_{ax}=p_{ax}$, which
have to satisfy four event-plane constraints given
in~\eqref{eq:softq}.  Then, introducing the unity,
\[
   1= \int dq_{1y} \prod_a dq_{ax}\cdot S_n\>,
\]
where  
\begin{equation}
\label{eq:Sdef}
  S_n\equiv \delta\!\left(q_{1y}\!+\!  \sum_R k_{iy}\right)
  \delta\!\left(q_{1x}\!+\!  \sum_R k_{ix}\right)
  \delta\!\left(q_{2x}\!+\!q^+_{1x}\!+\!\sum_U k_{ix}\right)
  \delta\!\left(q_{3x}\!+\!q^-_{1x}\!+\!\sum_D k_{ix}\right),
\end{equation}
we neglect hard parton recoils $q_{ai}$ in all but these four
components and approximate the kinematical factor $D_n$ as follows
\begin{equation}
  \label{eq:Dn'}
  \begin{split}
   D_n\> =\>\int dq_{1y} \prod_a dq_{ax}\> S_n\cdot D_n
\>\simeq \> D_0 \cdot \int dq_{1y} \prod_a dq_{ax}\cdot S_n\,.
  \end{split}
\end{equation}
Here $D_0$ is a trivial phase space factor which corresponds to the
Born three-parton kinematics,
\[
D_0=(2\pi)^4
\delta^4\left(\sum_a p_a -Q\right)
\delta^3\left(\vec{p}_1 -\vec{P}_1\right)
\delta^2\left(\vec{p}_{2t} -\vec{P}_{2t}\right).
\]  
We then have 
\begin{equation}
\label{eq:dGn}
d\Gamma_n \simeq \Gamma_0 \> \prod_{i=1}^n [dk_i]\,  dh_n\>, 
\qquad dh_n= dq_{1y} \prod_{a=1}^3 dq_{ax}\cdot S_n \,, \qquad
\Gamma_0= \prod_{a=1}^3 \> \frac{d^3p_a}{(2\pi)^32E_a}\cdot D_0\,, 
\end{equation}
where $\Gamma_0$ is the Born phase space given in Appendix~\ref{App:kin}.

Neglecting bremsstrahlung in the $D_0$-factor proves to be legitimate:
it can be shown that to achieve SL accuracy it suffices to take care
of accompanying parton momenta in the $S$-factor \eqref{eq:Sdef} and
in the observable itself.

Finally, in order to ``exponentiate'' the multiple radiation we need
to factorize dependence on individual secondary parton momenta
contained in the delta-functions in $S_n$ and in the $\Ko$-observable.
This is achieved in a standard way by means of Mellin and Fourier
representations.  In the following we apply this procedure to the
integrated $\Ko$-distribution defined in \eqref{eq:Sig}.

In what follows we shall separately consider $\Ko$ accumulated in the
right hemisphere and the total $\Ko$ of the event.  We start from a
simpler case of the $\Ko$-distribution in the {\em right}\/ (one-jet)
hemisphere. The total $\Ko$-distributions will be considered in
Sec.~\ref{sec:tot}.

\section{Right $\Ko$-distribution\label{sec:rig}}
Consider the differential three-jet cross section with given $T$ and
$T_M$ and with accumulated out-of-event-plane momentum in the right
hemisphere smaller than a given $\Ko$.  Using the soft-factorization
formula~\eqref{eq:Mn2} we can write the cross section \eqref{eq:Sig}
for small $\Ko$ in the form
\begin{equation}
  \begin{split}
\label{eq:SigfactR}
\frac{d\sigma^R(\Ko)}{dTdT_{M}} 
&=\sum_n \frac{1}{n!}\int M^2_{n}d\Gamma_n
\vartheta\left(\!\Ko\!-\!\abs{q_{1x}}\!-\!
\sum_R\abs{k_{ix}}\!\right)\\
&=\sum_{\conf=1}^3 \frac{d\sigma_{\conf}^{(0)}}{dT dT_M} 
\cdot \Sigma_{\conf}^R(\Ko)\>,
\end{split}
\end{equation}
where $\sigma_{\conf}^{(0)}$ is the three-jet differential Born cross
section for the parton configuration $\cC_{\conf}$ defined in
\eqref{eq:conf}, 
\[
\frac{d\sigma_{\conf}^{(0)}}{dT dT_M} \equiv \Gamma_0 M_{0}^2(\cC_{\conf})\>, 
\]
calculated in Appendix~\ref{App:kin}.  The accompanying radiation
factor $\Sigma$, a function of $\Ko$, $T$ and $T_M$, reads
\begin{equation}
\label{eq:SigR}
\begin{split}
&
\Sigma_{\conf}^R(\Ko)
=\int \>\sum_n \frac1{n!} \prod_{i}^n [dk_i]\> 
W_{\conf}(k_i)\cdot H^R(\Ko)\>,\\
& H^R(\Ko)\>\equiv
\>\int dh_n
\vartheta\left(\!\Ko\!-\!\abs{q_{1x}}\!-\!
\sum_R\abs{k_{ix}}\!\right).
\end{split}
\end{equation}
We recall that within the adopted PT accuracy the hard parton momenta
$p_a$ entering into the soft gluon distribution factor $W_{\conf}(k)$
can be approximated by the event plane vectors $P_a$ so that
integration over the recoil variables $q_a=p_a-P_a$ can be easily
performed.
Since the observable involves only momenta in the right hemisphere,
the recoil momentum components in the left hemisphere $q_{2x}$ and
$q_{3x}$ can be freely integrated out with use of the last two
delta-functions in \eqref{eq:Sdef} and leave no trace in the
distribution under consideration.  Then, a non-trivial dependence on
$q_{1y}$ (via $q_{1x}^\pm)$) also disappears, and the
$q_{1y}$--integration trivializes as well.  We are left with a single
delta-function in \eqref{eq:Sdef}:
\[
H^R(\Ko)\>= \>\int dq_{1x}\>\vartheta\left(\!\Ko\!-\!\abs{q_{1x}}\!-\!
\sum_R\abs{k_{ix}}\!\right)\,
\delta\left(q_{1x}+\sum_R k_{ix}\right)\>.
\]
To factorize the dependence on secondary parton momenta we use the
Mellin representation for the theta-function in $\Ko$ and the Fourier
representation for the delta-function in $q_{1x}$:
\begin{eqnarray}
\label{eq:nudef}
 \vartheta\left(\!\Ko\!-\!\abs{q_{1x}}\!-\!
\sum_R\abs{k_{ix}}\!\right) &=& \int_{{\cal C}} \frac{d\nu}{2\pi i\, \nu} 
\exp\left\{\nu\cdot\left(\!\Ko\!-\!\abs{q_{1x}}\!-\!
\sum_R\abs{k_{ix}}\!\right)\right\} , \\
\label{eq:bedef}
\delta\left(\!q_{1x}\!+ \sum_R {k_{ix}}\! \right) &=& 
\nu \int_{-\infty}^{\infty}
\frac{d\be}{2\pi}
\exp\left\{-i\nu\,\be\cdot \left(\!q_{1x}\!+ \sum_R {k_{ix}}\!\right)\right\},
\end{eqnarray}
where the contour ${\cal C}$ in \eqref{eq:nudef} runs parallel to the
imaginary axis at $\Real \> \nu>0$.  Defining the Fourier integral
\eqref{eq:bedef} we have extracted, for the sake of convenience, the
factor $\nu$ {\em as if}\/ it were a real parameter: for a complex
value of $\nu$ it implies rotating the $\be$-contour by $-\Arg\nu$,
$\abs{\Arg\nu}<\pi/2$. (Then, using the analytic continuation, the
$\be$-integral can be transformed to run along the real axis.)

Integrating over $q_{1x}$ we obtain
\begin{equation}
\label{eq:HR}
H^R(\Ko)\>=\>\int_{{\cal C}} \frac{d\nu}{2\pi i \nu}\>e^{\nu \Ko}
\int_{-\infty}^{\infty} \frac{d\be}{\pi(1+\be^2)}
\cdot \prod_R e^{-\nu (|k_{ix}| + i\be k_{ix})}\>.
\end{equation}
The limit of small $\Ko$ (and thus of small $q_{1x}$) corresponds to
the region of large values of the conjugate variables, i.e.\ $\nu$ and
$\nu\be$ respectively.  Therefore in what follows we will concentrate
on the limit of large~$\nu$ and neglect the contributions of the order
of $\nu^{-1}$ which correspond to $\cO{\Ko}$ corrections to the
distribution.
At the same time, by examining \eqref{eq:HR} it is easy to see that
the characteristic values of the rescaled Fourier variable $\be$ are
of the order of unity.

Substituting \eqref{eq:HR} into \eqref{eq:SigR} we get
\begin{equation}
  \label{eq:SigR1}
 \Sigma_{\conf}^R(\Ko)= \int_{{\cal C}} \frac{d\nu}{2\pi i \nu}\>e^{\nu \Ko}
\int_{-\infty}^{\infty} \frac{d\be}{\pi(1+\be^2)}
e^{-\cR_{\conf}^R(\nu,\be)}\>,
\end{equation}
where we have introduced the ``radiator''
\begin{equation}
  \label{eq:Rad}
  \cR_{\conf}^R(\nu,\be)=\int_R [dk]\>W_{\conf}(k)\>
\left[\>1\>-\>e^{-\nu(|k_x|+i\be k_x)}\>\right].
\end{equation}
Here the unity in the square brackets has been included to account for
the virtual corrections.  

For example, for the most probable jet configuration $\conf=3$ (with
gluon the least energetic parton), the soft distribution \eqref{eq:W3}
gives
\begin{equation}
  \label{eq:Rada}
  \cR_3^R=\frac{N_c}{2}
\left(r^R_{13}+r^R_{23}-\frac{1}{N_c^2}r^R_{12}\right),
\qquad
r^R_{ab}=\int_R [dk]\>w_{ab}(k)\>
\left[\>1\>-\>e^{-\nu(|k_x|+i\be k_x)}\>\right].
\end{equation}
With the DL accuracy only gluons collinear to $P_1$ (the thrust axis
direction) contribute to the right-hemisphere $\Ko$.
Therefore the (identical) DL contributions are contained in $r^R_{12}$
and $r^R_{13}$. According to \eqref{eq:Rada}, they combine into the
expression proportional to $C_F$ --- the colour charge of the quark
$P_1$.  Single logarithmic corrections to these two dipoles, as well
as SL contribution of the third dipole, $r^R_{23}$, are calculated in
Appendix~\ref{App:RadR}.  Here we report the result:
\begin{equation}
  \label{eq:R1a}
\begin{split}
r^R_{1a}=r(\bmu,Q^2)+
F_{1a}\int_{1/\bmu}^Q\frac{dk_x}{k_x}\frac{\as(2k_x)}{\pi}\>,
\qquad
r^R_{23}=
F_{23}\int_{1/\bmu}^Q\frac{dk_x}{k_x}\frac{\as(2k_x)}{\pi}\>,
\end{split}
\end{equation}
where the variable $\bmu$ originates from an approximate evaluation of
the characteristic momentum integral which is explained in
Appendix~\ref{App:theta},
\begin{equation}
  \label{eq:theta}
\left[\>1\>-\>e^{\nu(|k_x|+i\be k_x)}\>\right]
\>\to\>
\theta\left(|k_x|-\frac{1}{\bmu}\right), 
\qquad \bmu = \bnu \mu\,,\quad \mu=\sqrt{1+\be^2},
\end{equation}
with
\begin{equation}
  \label{eq:bnu}
\bnu\equiv \nu e^{\gamma_E}.
\end{equation}
In \eqref{eq:R1a} $r(\bmu,Q^2)$ is the DL function
\begin{equation}
  \label{eq:r}
  r(\bmu,Q^{'2})=\int_{1/\bmu}^{Q'}\frac{dk_x}{k_x}\frac{\as(2k_x)}{\pi}
\ln\frac{Q^{'2}}{k^2_x}\>,
\end{equation}
and the factors $F_{ab}=F_{ab}(T,T_M)$ are independent of the
integration variables $\nu$ and $\be$.
The origin of an essential subleading correction embodied into the
precise argument of the running coupling, $\as(2k_x)$, is explained in
Appendix~\ref{App:running}.

Combining these results we obtain the radiators, evaluated with SL
accuracy,  for each of the three kinematical jet configurations,
\begin{equation}
  \label{eq:R1a'}
     \cR^R_{\conf}(\bmu) =C_{\conf}\>r(\bmu,Q^2_{\conf})\>.
\end{equation}
Here $C_{\conf}$ is the colour charge of the hard parton along the
thrust axis, that is $C_{\conf}=C_F$ for $\conf=2,3$ and $C_1=N_c$.
SL corrections in \eqref{eq:R1a'} were absorbed into the definition of
the hard scale. The corresponding scale $Q_{\conf}$ depends on the
event configuration $\conf$ and, though the functions $F_{ab}(T,T_M)$,
on the event kinematics.  
These scales are given by
\begin{equation}
  \label{eq:scaleR}
\begin{split}
&  N_C\>\ln Q^2_1=N_C\>\ln (Q^2 e^{-\frac{\be_0}{2N_c}})
+\frac{N_c}{2}\>(F_{12}+F_{13}-\frac{1}{N_c^2}F_{23})\,,\\
&  C_F\>\ln Q^2_2=C_F\>\ln (Q^2 e^{-\frac{3}{2}})
+\frac{N_c}{2}\>(F_{12}+F_{23}-\frac{1}{N_c^2}F_{13})\,,\\
&  C_F\>\ln Q^2_3=C_F\>\ln (Q^2 e^{-\frac{3}{2}})
+\frac{N_c}{2}\>(F_{13}+F_{23}-\frac{1}{N_c^2}F_{12})\,.
\end{split}
\end{equation}
They also include the factors $e^{-3/4}$ and $e^{-\be_0/4N_c}$ coming
from SL corrections due to the ``hard'' parts of the quark and gluon
splitting functions.
As we shall see below in Sec.~\ref{sec:tot}, for the case of the {\em
  total}\/ $\Ko$-distribution the corresponding scales are simply
related with geometry of the 3-jet event.  At the same time, the $T$-
and $T_M$-dependence of the scales entering into one-hemisphere
distribution does not have a simple geometrical interpretation.  This
is due to the fact that the kinematical constraint restricting the
observable to a single hemisphere is foreign to the structure of the
soft-gluon radiation pattern.

The radiator \eqref{eq:R1a'} depends on the Mellin-Fourier moments
$\nu$ and $\be$ only via the variable $\bmu$.  The $\be$-dependence
can be further simplified at SL accuracy by expanding the radiator for
large~$\nu$,
\begin{equation*}
  \label{eq:expR}
  \cR^R_{\conf}(\bmu)=\cR^R_{\conf}(\bnu)+
C_{\conf}\>r'(\bnu)\>\ln\sqrt{1+\be^2}\>,
\end{equation*}
where
\begin{equation}  \label{eq:r'}
  r'(\bnu)=\frac{\as(
k_x)}{\pi}\ln \frac{Q^2}{k^2_x}\>,\qquad
k_x\equiv 1/\bnu\>.
\end{equation}
Since $r'$ constitutes a SL correction, we can use $Q$ as a common
scale (neglecting $\cO{\as}$ mismatch) and omit the factor $2$ in the
running coupling argument (as producing a negligible correction
$\cO{\as^2\log\bnu}$) in such subleading terms.

Thus, the right-hemisphere $\Ko$-distribution to SL accuracy takes the
form
\begin{equation}
  \label{eq:SigRfin}
\begin{split}
& \Sigma_{\conf}^R(\Ko)= \int \frac{d\nu}{2\pi i \nu}\>e^{\nu \Ko}
  \>e^{-\cR_{\conf}^R(\bnu)}\>\cF^R_{\conf}(\bnu)\>, \\
&
  \cF^R_{\conf}(\bnu)=\int_{-\infty}^{\infty} 
  \frac{d\be}{\pi(1+\be^2)^{1+\eta}}\>
= \frac{\Gamma(\half+\eta)}{\sqrt{\pi}\Gamma(1+\eta)}\>,
  \qquad \eta=\half C_{\conf}\>r'(\bnu)\>.
\end{split}
\end{equation}
Integration over the Mellin variable $\nu$ can be performed by
steepest descent or by the operator method introduced in
\cite{broadNP}.  We follow the last method which exploits the
following identities:
\begin{equation*}
  \label{eq:Op}
\begin{split}
 f(\bnu)\>=&\>\left.  
 f(e^{-\partial_z})\cdot\big(\bnu\big)^{-z}\right|_{z=0}\>; \\
  \int\frac{d\nu}{2\pi i \nu}\> \big(\bnu\big)^{-z} \>e^{\nu\Ko}\>
=&\> \frac{(\bKo)^z}{\Gamma(1+z)} \>,\qquad   
\bKo \>\equiv \>  e^{-\gamma_E}\Ko\>.
\end{split}
\end{equation*}                                
The following approximation is valid:
\begin{equation*}
\label{eq:op2}
 \left. e^{-R(e^{-\partial_z})}\cF(e^{-\partial_z}) 
 \frac{(\bKo)^z}{\Gamma(1+z)}\right|_{z=0}
 \simeq  e^{-R(\bKo^{-1})}\cF(\bKo^{-1})\cdot  
 \Gamma^{-1}\left(1+R'(\bKo^{-1})\right)\,,
\end{equation*}
where we have neglected relative corrections of the order
$R''(x)\equiv x\partial_x R'(x)=\cO{\as}$ and assumed that $\cF$ is a
smooth function of $R'$ which is true for $\cF^R_{\conf}(\bnu)$ in
\eqref{eq:SigRfin}, $d\ln\cF/d\eta=\cO{1}$.

Applying these relations to our distribution in \eqref{eq:SigRfin} we
derive the final answer to SL accuracy,
\begin{equation}
\label{eq:finR}
   \Sigma^R_{\conf}(\Ko)= e^{-\cR^R_{\conf}\left(\bKo^{-1}\right)} \cdot 
   \frac{\cF^R_{\conf}\left({\bKo}^{-1}\right)}
   {\Gamma\left(1+ C_{\conf} r'(\bKo^{-1})\right)}\>, \qquad
   \bKo \>=\>  e^{-\gamma_E}\Ko\>.
\end{equation}
The first and second factors resum DL and SL contributions,
respectively.  We remark that the precise argument $\bKo$ is essential
to keep in the first factor, while substituting $\bKo$ by, say, $\Ko$
in the second factor would amount to a negligible $\cO{\as}$
correction.
Comparing the $\nu$-integrand in \eqref{eq:SigRfin} with the final
result we conclude that the SL factor $\Gamma^{-1}$ in \eqref{eq:finR}
accounts for a mismatch between the Mellin-conjugated $\nu$ and $\Ko$
values.  It can be looked upon as a next-to-leading order prefactor of
the WKB (steepest descent) approximation.

\section{Total $\Ko$-distribution\label{sec:tot}}
The calculation is similar to the previous case except that now
radiation in all four quadrants contributes to $\Ko$.  The
$\Ko$-integrated distribution \eqref{eq:Sig} for small $\Ko$ is given
by
\begin{equation}
  \begin{split}
\label{eq:Sigfact}
&\frac{d\sigma(\Ko)}{dTdT_{M}} 
=\sum_{\cC_{\conf}} \frac{d\sigma_{\conf}^{(0)}}{dT dT_M} 
\cdot \Sigma^T_{\conf}(\Ko)
\>,\qquad \\
&\Sigma^T_{\conf}(\Ko)
=\int \>\sum_n \frac1{n!} \prod_{i}^n [dk_i]\> W_{\conf}(k_i)
\cdot H^T(\Ko)\>,\\
& H^T(\Ko)\>\equiv \>\int dh_n \vartheta\left(\!\Ko\!-\!\sum_{a=1}^3 
\abs{q_{ax}}\!-\! \sum_i\abs{k_{ix}}\!\right).
\end{split}
\end{equation}
To factorize the soft momenta in $H^T$ we proceed as before by using
Mellin and Fourier representation for the theta- and delta-functions.
Again we denote by $\nu$ the variable conjugate to $\Ko$ and study the
region $\abs{\nu}Q\gg1$.  We rescale by $\nu$ the variables conjugate
to the soft recoil variables $q_{ax}$ ($a=1,2,3$) and $q_{1y}$ in
$dh_n$ to arrive at
\begin{equation}
  \label{eq:hn}
  \begin{split}
&\int dh_n
\prod_{a=1}^3 e^{-\nu\abs{\!q_{ax}\!}}
\prod_{i=1}^n e^{-\nu\abs{\!k_{ix}\!}}\>=\>\\
&\int_{-\infty}^{\infty} \frac{d\gamma}{2\pi}
\left(
\prod_{a=1}^3\! \frac{d\be_{a}}{\pi}
\right)      
\>I(\be,\gamma) \left\{
\prod_{C_1}u_{12}(k_i)\prod_{C_4}u_{13}(k_i)
\prod_{C_2}u_2(k_i)\prod_{C_3}u_3(k_i)\right\}.
\end{split}   
\end{equation}
Here we have introduced the probing functions $u_\alpha(k)$ for each
of the quadrants $C_{\ell}$:
\begin{equation}
  \label{eq:u}
\begin{split}
  &u_{12}(k)=u(\be_{12},\gamma)\>,\quad
   u_{13}(k)=u(\be_{13},-\gamma)\>,\quad
   u_{2}(k) =u(\be_{2},0)\>,\quad
   u_{3}(k) =u(\be_{3},0)\>,
 \end{split}
\end{equation}
where $\be_{12}=\be_1+\be_2$,  $\be_{13}=\be_1+\be_3$ and
the ``source function'' $u$ is
\begin{equation}
  \label{eq:u'}
\begin{split}
  u(\be,\gamma)\equiv 
\exp\left\{-\nu\left(\abs{\!k_{ix}\!}+i\be k_{ix}+i\gamma |k_{iy}|
\right)\right\}.
\end{split}
\end{equation}
The function $I(\be,\gamma)$ in \eqref{eq:hn} is the result of
integrations over the recoil momenta, namely, $q_{1y}$ (conjugate to
$\gamma$) and three $q_{ax}$ (conjugate to $\be_a$):
\begin{equation}
\begin{split}
\label{eq:I}
I(\be,\gamma)
=\frac{1}{1+\be_2^2}
 \>\frac{1}{1+\be_3^2} 
 \left(  \frac{1}{1+\be_{12}^2}\>\frac{1}{-i\gamma+\eps} + 
         \frac{1}{1+\be_{13}^2}\>\frac{1}{ i\gamma+\eps} 
\right).
\end{split}
\end{equation}
Notice that the large variable $\nu$ enters only in the sources.  

Now we are in a position to resum multiple accompanying radiation.
The result reads
\begin{equation}
\begin{split}
\label{eq:sf}
\Sigma_{\conf}(\Ko) = \int \frac{d\nu}{2\pi i\nu} \> e^{\nu\Ko} \>
\int \frac{d\gamma}{2\pi} \> \prod _{a=1}^3\frac{ d\be_a}{\pi} 
\>I(\be,\gamma)
\cdot  e^{-\cR_{\conf}(\nu,\be,\gamma)}\>,
\end{split}
\end{equation}
with  the radiator given by
\begin{equation}
  \label{eq:RadTot}
\begin{split}
  \cR_{\conf}(\nu,\be,\gamma)=\int [dk]\>W_{\conf}(k)\>
\Big[1-\sum_{\ell=1}^4 u_{\ell}(k)\,\Theta_{\ell}(k) \Big]\>.
\end{split}
\end{equation}
As before, the unity in the square brackets has been included to
account for the virtual correction contribution.  Here
$\Theta_{\ell}(k)$ is the support function for a parton $k$ emitted in
the quadrant $C_{\ell}$, and we have denoted $ u_1=u_{12}$ and
$u_4=u_{13}$.

In Appendix~\ref{App:Rad} the radiators $\cR_{\conf}$ are evaluated
with SL accuracy and we obtain
\begin{equation}
\label{eq:RadTT}
\begin{split}
\cR_\conf = 
C_2^{(\conf)}\>r(\bmu_2,Q_2^2)+
C_3^{(\conf)}\>r(\bmu_3,Q_3^2)+
\frac{C_1^{(\conf)}}{\pi}\int_0^\infty \frac{dy}{1+y^2}
\>\left[\,r(\bmu_{12},Q_1^2) + r(\bmu_{13},Q_1^2)\,\right],
\end{split}
\end{equation}
where $r$ is the DL function defined in \eqref{eq:r}.  Some comments
on the colour charges, the scales and the various $\bmu$-variables are
in order.
\begin{itemize}
\item 
  The radiator consists of three ``independent'' contributions from
  the radiation off each of three hard partons. Here 
\begin{equation}
  \label{eq:Cconf}
C^{(a)}_a=N_c\,; \qquad   C^{(b)}_a=C_F\,, \quad \mbox{for}\> a\ne b\,,
\end{equation}
is the colour charge of the parton $P_a$.
\item 
  The hardness scale $Q_a^2$ in each term has a simple structure: it
  is determined by the invariant transverse momentum of the hard
  parton $a$ with respect to the dipole $bc$:
\begin{equation}
\label{eq:scalesT}
\begin{split}
Q_{a}^2 =  \frac{p_{ta}^2}{4}e^{-g_a}\>, \qquad
p_{ta}^2=  2\>\frac{(P_bP_a)(P_aP_c)}{(P_bP_c)}\>.
\end{split}
\end{equation}
This makes $Q_a$ depending on the event geometry.  The scale $Q_a^2$
also includes an additional, geometry-independent, factor $e^{-g_a}$
depending on the nature of the parton $a$. This factor takes into
account a SL correction due to hard parton splitting, with $g_a=3/2$
for a quark ($a=1,2$ in \eqref{eq:RadTT}) and $g_a=\be_0/2N_c$ for a
gluon. Due to this factor the scales depend on the configuration
$\conf$.

Recall that the derivation of the exact form of the hard scales
required a SL analysis which takes a due care of the inter-jet regions
where the soft distribution does not have collinear singularities.

As we have seen in the previous section, the hard scales for the right
distribution \eqref{eq:scaleR} do not have such a simple
interpretation because selecting one hemisphere is unnatural for the
soft radiation pattern.  As shown in Appendices~\ref{App:d23} and
\ref{App:d1a}, the boundary effects due to radiation at 90$^o$ which
complicate the scales, cancel with SL accuracy in the {\em total}\/
distribution.

\item 
  The various functions $\bmu$ originate from the following
  substitutions in the integral over $k_x$
\begin{eqnarray}
\label{theta}
  1-u_{a} &\to& \vartheta(k_x-1/\bmu_{a})\>, \quad\>\>
   \bmu_{a}\> = \bnu\,\sqrt{1+\beta_{a}^2} 
   \equiv\bnu\cdot \mu_{a} \>, \quad a=2,3 \\ 
   1-u_{12} &\to& \vartheta(k_x-1/\bmu_{12})\>, \quad
   \bmu_{12} = \bnu\,  
          \sqrt{(1-i\gamma y)^2+\beta_{12}^2}
          \equiv\bnu\cdot \mu_{12}\>. \\ 
   1-u_{13} &\to& \vartheta(k_x-1/\bmu_{13})\>, \quad
          \bmu_{13} = \bnu\,
          \sqrt{(1+ i\gamma y)^2+\beta_{13}^2} 
          \equiv\bnu\cdot \mu_{13}\>,
\end{eqnarray}
with $\bnu$ given in \eqref{eq:bnu}.
As in the case of the right radiator, such substitution is valid
within SL accuracy.
The $\gam$-dependence enters only in the two $\bmu_{1a}$ parameters
which are associated with the gluon emission off the parton $P_1$ in the
right hemisphere.  This is in agreement with the fact that the
variable $\gam$ is conjugate to $q_{1y}$.
Opposite signs of $\gam$ in $\bmu_{12}$ and $\bmu_{13}$ reflect the
fact that the recoil momentum $q_{1y}$ is positive (negative) in the
right-up (right-down) quadrant.

All $\nu$-, $\be$- and $\gam$-dependence is contained in the
parameters $\bmu$.
\end{itemize}
As before, we can simplify the $\be$- and $\gam$-dependence by
expanding the various terms to SL accuracy for large $\nu$.  
We can write
\begin{equation}
  \label{eq:Radfin}
  \cR_\conf = R_\conf(\bnu) + r'(\bnu)\cdot S_\conf(\be,\gam)\>,
\end{equation}
where $R_\conf$ is the DL contribution, given by the sum of
three standard antenna terms, $r'$ is the SL function defined in
\eqref{eq:r'}, and the coefficient $S_\conf$ carries all the
$\be_a$ and $\gam$ dependence.  $R_\conf$ and $S_\conf$ are
given by
\begin{equation}
  \label{eq:Rconf}
\begin{split}  
&   R_\conf(\bnu) = C_1^{(\conf)}\,r(\bnu,Q_1^2) +
C_2^{(\conf)}\,r(\bnu,Q_2^2) +C_3^{(\conf)}\,r(\bnu,Q_3^2)\,,\\
&  S_\conf(\be,\gam)=
C_2^{(\conf)}\>\ln\mu_{2}\!+\!
C_3^{(\conf)}\>\ln\mu_{3}\!+\! 
\frac{C_1^{(\conf)}}{\pi}\int_0^\infty \frac{dy}{1+y^2} 
\ln\left[\,\mu_{12}(y)\mu_{13}(y)\,\right],
\end{split}
\end{equation}
with the charges given in \eqref{eq:Cconf}.
In detail the DL terms for the three configurations are 
\begin{eqnarray}
\label{eq:R33}
 R_3(\bnu) &=& C_F\,r(\bnu,Q_1^2) +C_F\,r(\bnu,Q_2^2) +N_c\,r(\bnu,Q_3^2)\,,\\
\label{eq:R32}
 R_2(\bnu) &=& C_F\,r(\bnu,Q_1^2) +N_c\,r(\bnu,Q_2^2) +C_F\,r(\bnu,Q_3^2)\,,\\
\label{eq:R31}
 R_1(\bnu) &=& N_c\,r(\bnu,Q_1^2) +C_F\,r(\bnu,Q_2^2) +C_F\,r(\bnu,Q_3^2)\,.
\end{eqnarray}
There are two sources of SL corrections in \eqref{eq:Radfin}.
The first is due to different hard scales in three terms
$r(\bnu,Q_{1})$ in \eqref{eq:R33}--\eqref{eq:R31}, which depend on
the geometry of the three-jet event, that is on the values of $T$ and
$T_M$.
The second is the contribution proportional to $r'$ given by the sum
of $\ln \mu$--terms which depend on $\be_a$ and $\gam$.  These
contributions are specific for a three-jet topology.  At the same
time, within the SL accuracy this correction is insensitive to the
details of the event geometry, that is to the values of $T,T_M\sim 1$.

In conclusion, the total $\Ko$-distribution, to SL accuracy, can be
expressed by the following Mellin integral:
\begin{equation*}
  \Sigma_{\conf}(\Ko) = 
\int\frac{d\nu}{2\pi i\nu}\> e^{\nu\Ko}\>
e^{-R_{\conf}(\bnu)}\cdot \cF_{\conf}(\bnu) \>,
\end{equation*}
where the 
SL prefactor
$\cF_{\conf}(\bnu)$ is given by 
\begin{equation}
\cF_{\conf}(\bnu) =\int_{-\infty}^\infty
\frac{d\gamma}{2\pi}\>\prod_{a=1}^3 \int_{-\infty}^\infty \frac{d\beta_a}{\pi}
\cdot I(\beta,\gamma)\>\>e^{-r'(\bnu)S_{\conf}(\be,\gam)}.
\end{equation}
For example, the explicit expression for $\cF_3$ reads 
\begin{equation*}
\begin{split}  
\cF_{3}(\bnu) =& 
\int\frac{d\gamma}{2\pi}\prod_{a=1}^3 \frac{d\beta_a}{\pi}
\cdot I(\beta,\gamma)
\>\left({1+\be_2^2} \right)^{-\half C_F r'(\bnu)}
\>\left({1+\be_3^2} \right)^{-\half N_c r'(\bnu)}
\\&\times
\exp\left\{ -C_F r'(\bnu) \left\{ 
\frac{1}{\pi}\int_0^\infty \frac{dy}{1+y^2} 
\ln\left[\,\mu_{12}(y)\mu_{13}(y)\,\right] \right\}\right\}.
\end{split}
\end{equation*}
The function $\cF_{\conf}$ is analysed 
in Appendix~\ref{App:cF}.  

In the first order in $\as\ln\nu$ the factors $\cF$ become 
(see Appendix~\ref{App:cF1})
\begin{equation*}
  \begin{split}
&\cF_{\conf}=1-  (3C_F+N_c)\>\ln 2\cdot r'(\bnu)\>,\qquad \conf=2,3\>,\\
&\cF_{1}=1- (2C_F+2N_c)\>\ln 2 \cdot r'(\bnu)\>.
\end{split}
\end{equation*}
Different weights of the quark and gluon colour factors for the two
cases have a simple explanation.  Due to the kinematics of parton
recoil (see \eqref{eq:softq}) contribution of the gluon radiation off
the most energetic (right-hemisphere) parton $P_1$ is {\em twice}\/
that off the left-hemisphere partons $P_2$ and $P_3$.  As a result,
the SL correction is proportional to $2C_F+C_F+N_c$ when $P_1$ is a
quark/antiquark ($\conf=2,3$) and to $2N_c+C_F+C_F$ when it is a gluon
($\conf=1$).

Integration over the Mellin variable $\nu$ can be done as before
and we obtain
\begin{equation}
\label{eq:StotT}
   \Sigma_{\conf}(\Ko) \simeq e^{-R_{\conf}\left(\bKo^{-1}\right)} \cdot 
\frac{\cF_{\conf}\left({\bKo}^{-1}\right)}
{\Gamma\left(1+R'(\bKo^{-1})\right)}\>,
\qquad  \bKo \>\equiv \>  e^{-\gamma_E}\Ko\>,
\end{equation}
where $R'$ is the logarithmic derivative of $R_{\conf}$. To SL
accuracy we can take
\begin{equation}
\label{eq:R'}
\qquad R'(\bKo^{-1})\>=\> (2C_F+N_c)\cdot r'(\bKo^{-1})\,,
\end{equation}
which is the same function for all configurations $\conf$. This is
possible since, $r'$ being a SL function, the difference between the
hard scales can be neglected at the level of the {\em
  next-to-next-to-leading}\/ $\cO{\as}$ correction.

\subsection{Radiators in the quasi-2-jet limit\label{sec:2jet}}
In the 3-jet kinematics we have been considering, $T_M\la T\la 1$, jet
energies are comparable and relative angles between jets are large. In
these circumstances three basic scales in \eqref{eq:scalesT} are of
the same order, $Q_1\sim Q_2\sim Q_3\la Q$.  Still, keeping precise
scales in \eqref{eq:R33}--\eqref{eq:R31} is essential, since their
deviation from the overall hardness parameter $Q$ in the DL radiator
function $r(\bnu,Q_a^2)$ produced a SL correction $\delta^{(1)}
R\propto r'(\bnu,Q)\cdot \ln(Q_a/Q)=\cO{\as\log\nu}$.  
At the same time, since the next order expansion terms are negligible,
$\delta^{(2)} R\propto r''(\bnu,Q)\cdot
\ln^2(Q_a/Q)=\cO{\as
%\ln^2(Q_a/Q)
}$, it is perfectly legitimate to present the answer in a form
different from \eqref{eq:R33}--\eqref{eq:R31}.

For example, expanding the scales in the first two (quark) terms in
$R_3$ around $q^2=2P_1P_2$ we obtain, instead of \eqref{eq:R33}
\begin{equation}
  \label{eq:R3exp}
  R_3 \>\simeq\> 2C_F\,r\left(\frac{P_1P_2}{2}e^{-\frac32}\right)
             \>+\>N_c\,r(Q_3^2)\>,\qquad  
Q_3^2=\frac{2(P_1P_3)(P_2P_3)}{(P_1P_2)}e^{-\frac{\be_0}{2N_c}}\,,
\end{equation}
where we have suppressed the first argument $\bnu$ and
used $\ln\frac{p_{t1}^2}{2P_1P_2}=-\ln\frac{p_{t2}^2}{2P_1P_2}$
to cancel the linear expansion terms, see \eqref{eq:scalesT}. 
Mismatch between the radiator in \eqref{eq:R33} and the right-hand
side of \eqref{eq:R3exp} is 
$$
R_3(\ref{eq:R3exp}) - R_3(\ref{eq:R33})\>=\>
 \cO{\as\ln^2\frac{P_1P_3}{P_2P_3}}\>.
$$
Being equivalent within the adopted accuracy, the representations
\eqref{eq:R33} and \eqref{eq:R3exp} start to significantly differ,
however, when the jet configuration becomes 2-jet-like.  Indeed, when
the gluon jet $P_3$ becomes relatively soft and/or collinear to the
quark $P_2$, we should expect the answer to correspond to the
$\qq$-dominated radiation pattern, and the gluon contribution to
disappear.

The latter representation \eqref{eq:R3exp} correctly describes this
situation: the quark-antiquark antenna contribution with the hardness scale
equal to the invariant squared mass of the $\qq$ pair, $q^2=2P_1P_2$,
takes on the job, while radiation off the gluon vanishes with decrease
of the gluon transverse momentum, $Q_3^2\simeq p_{t3}^2\ll q^2\simeq Q^2$.
At the same time, the first representation goes hay-wire. For example,
in the collinear limit, $P_3P_2\ll P_3P_1\la P_1P_2$, one of the quark
scales, $Q_2^2$, vanishes (together with $Q_3^2$) while the other
formally goes to infinity, $Q_1^2\gg Q^2$.  The geometric mean of the
quark scales, $2P_1P_2=\sqrt{p_{1t}^2p_{2t}^2}$, employed in
\eqref{eq:R3exp} cures this unphysical behaviour.

For two other jet configurations, with the gluon having an intermediate or
the largest of the three parton energies, analogous representations of 
the radiators read 

\vspace{-0.7cm}

\begin{eqnarray}
  \label{eq:R2exp}
 R_2 &\simeq& \!\! 2C_F\,r\left(\frac{P_1P_3}{2}e^{-\frac32}\right)
                     \>+\>N_c\,r(Q_2^2) \,, \qquad 
Q_2^2=\frac{2(P_1P_2)(P_2P_3)}{(P_1P_3)}e^{-\frac{\be_0}{2N_c}}\,; 
  \label{eq:R1exp}
\end{eqnarray}

\vspace{-1cm}

\begin{eqnarray}
 R_1 &\simeq& \!\! N_c\left[\,                              
   r\left(\frac{P_1P_2}{2}e^{-(\frac34+\frac{\be_0}{4N_c})}\right) 
 + r\left(\frac{P_1P_3}{2}e^{-(\frac34+\frac{\be_0}{4N_c})}\right)\,\right]
-\frac{1}{N_c}\,r\left(\frac{P_2P_3}{2}e^{-\frac32}\right) .  
\end{eqnarray}
These expressions, contrary to the original ones \eqref{eq:R32} and 
\eqref{eq:R31}, survive the limit of the left-hemisphere jets becoming 
quasi-collinear, $P_2P_3\ll P_1P_3\la P_1P_2$.
Indeed, the configuration $\conf=2$ is then identical, colour-wise, to 
 $\conf=3$, and \eqref{eq:R2exp} is dominated by the $\qq$ two-jet
 contribution, while the non-Abelian part vanishes with the gluon
 transverse momentum, $Q_2^2\simeq p_{2t}^2\ll Q^2$.
 A rare but interesting configuration $\conf=1$, where the gluon $P_1$
 in the right hemisphere is balanced by a quasi-collinear $\qq$ pair
 in the left hemisphere, see \eqref{eq:R1exp}, corresponds to the
 gluon-gluon system: radiation off the {\em colour-octet}\/ $\qq$ pair
 is dominated by large-angle coherent bremsstrahlung proportional to
 the gluon charge, $N_c$, with the colour-suppressed $1/N_c$
 correction term vanishing in the collinear limit.
 
 Thus, the radiators in the form of \eqref{eq:R33}--\eqref{eq:R31} are
 inapplicable in the quasi-two-jet kinematics.  An attentive reader
 could have noticed that the modified expressions
 \eqref{eq:R3exp}--\eqref{eq:R1exp}, though better behaved, cannot
 pretend to uniformly preserve the desired SL accuracy.  Accommodating
 correctly SL effects due to large-angle soft bremsstrahlung, these
 expressions fail, however, to properly account for ``hard'' SL
 corrections in the collinear limit.

For example, the right-hand side of \eqref{eq:R3exp} for the most
natural configuration $\conf=3$ has a perfect {\em soft}--gluon limit, 
when $2(P_1P_2)=Q^2$ becomes the proper 2-jet scale. At the same time, 
when $P_3$ remains energetic but collinear, $E_3\la E_2$,
$(P_2P_3)\to0$, the quark-jet scale in \eqref{eq:R3exp} remains 
smaller than the total annihilation energy, 
$2(P_1P_2)<2(P_1\cdot(P_2+P_3))=Q^2$. The mismatch amounts to a SL
correction $\cO{r'\cdot \ln x_2}$.  

\section{Discussion and conclusions}

In this paper we performed the all-order perturbative analysis of the
out-of-plane transverse momentum distributions in three-jet $e^+e^-$
annihilation events.  We considered the total $\Ko$ of the event and
$\Ko^R$ accumulated in one event-hemisphere which contains the most
energetic of three jets.
The perturbative expression for the integrated $\Ko$-distribution in
the right (one-jet) hemisphere is given in
\eqref{eq:r}--\eqref{eq:finR} and \eqref{eq:Fabr}.
The $\Ko$-total distribution is determined by expressions 
\eqref{eq:RadTT}--\eqref{eq:R'}.
These answers resum all double- and single-logarithmic contributions
to the exponent (the so-called ``radiator'') of the distribution in
the standard Mellin-Fourier parameter space.  DL and SL contributions
to the radiator can be formally represented, respectively, as series
$\as^n\log^{n+1}(Q/\Ko)$ and $\as^n\log^{n}(Q/\Ko)$ ($n\ge1$).

Throughout the analysis, we systematically neglected
next-to-next-to-leading terms in the radiator (radiator series
$\as^n\log^{n-1}\ln (Q/\Ko)$, $n\ge1$) as well as non-exponentiating
corrections of the relative order $\cO{\as}$.  The latter belong to
the coefficient function $1+c{\as}+\cO{\as^2}$. Its first coefficient,
$c$, is analysed in~\cite{BSZ} where the order $\as^2$ expansion of
the approximate resummed cross is compared with numerical calculation
based on the exact $\as^2$ matrix element.
Matching the resummed logarithmic expressions with the exact result 
is necessary for justifying exponentiation of the next-to-leading 
SL terms $\as\ln(Q/\Ko)$. 

Accompanying gluon radiation pattern follows the colour topology of
the underlying parton antenna. The corresponding patterns were known,
to SL accuracy, for two-parton --- $\qq$ and $gg$ --- sources (quark
and gluon form factors, 2-jet shapes). 
In the present paper we presented the first such analysis for 
3-parton ensembles. 

The $\qq g$--initiated events possess a rich colour structure which
determines secondary parton flows and makes them
event-geometry-dependent.
We have demonstrated that after taking into account SL effects due to
inter-jet gluon flows, the result can be cast as a sum of
quark-antiquark ($C_F$) and gluon ($N_c$) contributions with the
proper hardness scales depending on $T$ and $T_M$.

A significant dependence on event geometry is the 
key feature which singles out the $\Ko$ distribution among other
$e^+e^-$ event shape observables.

Soft gluon field components with small transverse momenta (``gluers''
with $k_t\ga \LQCD$) are believed to be responsible for hadronization.
This belief, known as hypothesis of the local parton-hadron duality,
has been verified in a number of experimental studies of various
features of multiple hadroproduction in hard processes (for review
see~\cite{KO}).
Moreover, in recent years a new theoretical techniques have been
developed for triggering and quantifying genuine confinement
effects by studying gluer radiation. 

A rich dependence of gluon (and, therefore, gluer) radiation on the
colour topology makes 3-jet observables, and $\Ko$ in particular, an
interesting field for the study of non-perturbative effects.
A separate publication~\cite{KoNP} will be devoted to analysis of the
leading non-perturbative power-suppressed corrections to 3-jet
observables. The $\Ko$ distribution, similar to the case of 2-jet
Broadening observables~\cite{broadNP}, will be shown to exhibit
$\log\Ko$--enhanced $1/Q$ contribution with the magnitude depending on 
3-jet geometry, that is on $T$ and $T_M$ values. 

To access this interesting physics experimental studies of $\Ko$
should be carried out specifically for events with moderate values of
$1-T$, away from the 2-jet region. 

Two sorts of experimental studies of these predictions can be
envisaged. The most straightforward comparison calls for experimental
identification of the gluon jet.  Gluon tagging, however, is
unnecessary for the study of the total $\Ko$-distribution for genuine
3-jet events: the prediction given by the sum of three jet
configurations corresponding to given $T$ and $T_M$, weighted with the
proper 3-jet Born cross section factors bears practically as much
information.
Tagging brings in more information when a single-jet
(right-jet) $\Ko$-distribution is studied. In this case essentially
different $\Ko$-spectra will be seen depending on whether the
right-hemisphere parton is a quark or a gluon.

\section*{Acknowledgements}
We are grateful to Gavin Salam for helpful discussions and
suggestions.

\appendix
\section{Kinematics \label{App:kin}}
\subsection{Kinematics of $\qq g$ jets}
We consider the quark, antiquark and gluon with skeleton momenta $P_a$
in the centre-of-mass frame, $\sum_{a=1}^3 P_a=Q=(Q,0,0,0)$,
\begin{equation}
  \label{eq:Ps}
   P_1=E(x_1,0,0,t_1)\>,\quad
   P_2=E(x_2,0, T_M,-t_{2})\>,\quad
   P_3=E(x_3,0,-T_M,-t_{3})\>.  
\end{equation}
We have $2E=Q$, $x_a=2(P_aQ)/Q^2$,
$x_1+x_2+x_3=2$ and $t_1=t_2+t_3$.  Assuming $x_2 > x_3$, we obtain
\begin{equation}
  \begin{split}
  &x_1=T\>, \qquad \qquad \qquad \>\>
   t_1=T\>,\\ 
  &x_2=\frac{2-T}{2} +\frac{T}{2}\rho\>,\qquad 
   t_2=\frac{T}{2} +\frac{2-T}{2}\rho\>,\\ 
  &x_3=\frac{2-T}{2} -\frac{T}{2}\rho\>,\qquad
   t_3=\frac{T}{2} -\frac{2-T}{2}\rho\>,
  \end{split}
\end{equation}
where
\begin{equation}
  \label{eq:rhodef}
\rho\equiv\sqrt{1-\frac{T_M^2}{1-T}} < 1\,.  
\end{equation}
Thrust-major is confined to the kinematical region
\begin{equation}
  \label{eq:Tmlim}
\frac{2\>(1-T)}{T}\sqrt{2T-1}\> < \> T_M\><\> \sqrt{1-T}\>,  
\end{equation}
where the upper limit comes from reality of $\rho$, and the lower
limit comes from requiring $x_1>x_2$. It is straightforward to show that in
this kinematical region $t_3>0$, i.e.\ that both $P_2$ and $P_3$
lie in the left hemisphere.

We introduce Sudakov variables based on two light-light vectors
aligned with the thrust axis,
\begin{equation}
  \label{eq:PSud}
  P=E(1,0,0,1)\>,\qquad \bar P=E(1,0,0,-1)\>.
\end{equation}
The jet momenta have the following Sudakov decomposition:
\begin{equation}
 P_1=T P\>,\quad
 P_2=A_2 P+B_2\bar P +P_{t}\>,\quad
 P_3=A_3 P+B_3\bar P -P_{t}\>,
\end{equation}
where the transverse momentum with respect to the thrust axis is
$P_t=(0,0,ET_M,0)$. In terms of $T$ and $T_M$ the longitudinal
Sudakov momentum components are
\begin{equation}
  \begin{split}
&A_2=\frac{1-T}{2}(1-\rho)\>,\qquad
 B_2=\frac{1}{2}(1+\rho)\>,\\
&A_3=\frac{1-T}{2}(1+\rho)\>,\qquad
 B_3=\frac{1}{2}(1-\rho)\>.
  \end{split}
\end{equation}
Since $P_a$ with $a=2,3$ belong to the left hemisphere, we have $A_a<B_a$.

Denoting by $\Theta_a$ the angle of $\vec{P}_a$ with respect to the thrust axis
we introduce the angular variables $\spi_a$ 
\begin{equation}
\label{eq:spi}
  \begin{split}
&\frac{\spi_2}{Q}\equiv\tan\frac{\Theta_2}{2}
                 =\frac{P_t}{Q A_2}=\frac{1+\rho}{T_M}\>,\\
&\frac{\spi_3}{Q}\equiv\tan\frac{\Theta_3}{2}
                 =\frac{-P_t}{Q A_3}=-\frac{1-\rho}{T_M}\>.
  \end{split}             
\end{equation}
We remark that $\spi_3$ is negative since $P_3$ lies in the third
quadrant (``down'' hemisphere).  
Since $P_2$ and $P_3$ are in the left hemisphere and
$\abs{P_{2z}}>\abs{P_{3z}}$, we have
$0<\pi-\Theta_2<\Theta_3-\pi<\half\pi$. 
Hence,
\[
\frac{\spi_2}{Q} > \frac{-\spi_3}{Q} > 1\,. 
\]
The following relations hold:
\begin{equation}
  \label{eq:spispi}
\spi_2\abs{\spi_3}=\frac{Q^2}{1-T}\>,\qquad
\spi_2+\abs{\spi_3}=\frac{2Q}{T_M}\>.
\end{equation}
The minimal value of thrust-major, 
$T_M=2(1-T)/(2-T)$ for a given $T$ in \eqref{eq:Tmlim},
corresponds to the configuration with the softest parton momentum
orthogonal to the thrust axis, 
\[
\rho\to \frac{T}{2-T}\>, \qquad
\frac{\spi_2}{Q}\>\to\> \frac{1}{1-T}\>,\quad
\frac{\abs{\spi_3}}{Q}\>\to\> 1\,.
\]
The maximal value $T_M = \sqrt{1-T}$ is achieved with a symmetric
$2--3$ pair:
\[ 
\rho\to0\>, \qquad
\frac{\spi_2}{Q}\to \frac{\abs{\spi_3}}{Q} \>\to\> \frac{1}{\sqrt{1-T}}\,.
\]
We also introduce two-dimensional vectors in the transverse plane $\{x,y\}$,
\begin{equation}
\label{eq:spidef}
  \begin{split}
\vpi_2\equiv\frac{\vec{P}_t}{A_2}\>\>= (0,\spi_2)\>,
\qquad
\vpi_3\equiv\frac{-\vec{P}_t}{A_3}=(0,\spi_3)\>.
  \end{split}
\end{equation}
Since $P_1$ is along the thrust axis we have $\vpi_1=(0,0)$.

\subsection{Born cross sections}
The squared Born matrix element $M_{0}(\cC_{\conf})$ has a well-known
expression in terms of the variables $x_a$.  For $\conf=3$ (with $P_3$
the gluon momentum) we have
\begin{equation}
  M_{0}^2(\cC_3) = \frac{C_F}{2\pi} \as(Q) \>
\frac{x_1^2+x_2^2}{(1-x_1)(1-x_2)}\>.
\end{equation}
The phase space factor $\Gamma_0$ for the Born 3-parton system is 
\begin{equation}
  \label{eq:G0}
\begin{split}
  \Gamma_0&=\frac{1}{8(2\pi)^5}\prod_{a=1}^3\frac{d^3p_a}{E_a}
\delta^4(Q-\sum p_a)\delta^3(\vec{p}_1-\vec{P}_1)
\delta^2(\vec{p}_{t2}-\vec{P}_{t2})
\end{split}
\end{equation}
The dipole invariant masses in terms of $T,T_M$ and variables $\spi_i$ 
are
\begin{equation}
  \label{eq:Qa}
\begin{split}
&
  Q^2_{12}=2P_1P_2=\frac{T}{2}(1+\rho)Q^2=\half{TT_M \tau_2}Q\>,\quad
\\&
  Q^2_{13}=2P_1P_3=\frac{T}{2}(1-\rho)Q^2=\half{TT_M|\tau_3|}Q\>,\quad
\\&
  Q^2_{23}=2P_2P_3=(1-T)Q^2=\frac{Q^4}{\tau_2|\tau_3|}\>.
\end{split}
\end{equation}
The scales in \eqref{eq:scalesT},
\[
p^2_{ta}\>\equiv \> \frac{2(P_bP_a)(P_aP_c)}{(P_bP_c)}\,,
\]
--- the invariant transverse momentum of the hard parton $P_a$ with
respect to the $bc$-dipole --- read
\begin{equation}
  \label{eq:pta}
  p^2_{t1}=\frac{Q^2}{2}\left(\frac{TT_M}{1-T}\right)^2,\quad
  p^2_{t2}=Q^2(1-T)\left(\frac{1+\rho}{T_M}\right)^2,\quad
  p^2_{t3}=Q^2(1-T)\left(\frac{1-\rho}{T_M}\right)^2 .
\end{equation}

\subsection{Soft partons}
For secondary massless parton of momentum $k$ 
we write the Sudakov representation
\begin{equation} k=\al P + \be \bar P + k_t\>,\qquad
  \al\be=\frac{k_t^2}{Q^2}\>,
\end{equation}
with $\vec{k}_t$  the transverse momentum with respect to the
thrust axis.
The right-hemisphere condition imposes the restriction upon 
longitudinal parton components, 
\[
\al>\be\>\>\Rightarrow\>\>\al>\frac{k_t}{Q}\>.
\]
Probability of soft gluon emission  off the $P_a,P_b$ dipole is
described by the squared matrix element 
\begin{equation}
\label{eq:dip}
\frac{P_aP_b}{2(P_ak)(kP_b)}= \frac{\al^2\left(\vpi_a-\vpi_b\right)^2}
{\left(\vec{k}_t-\al\vpi_a\right)^2\left(\vec{k}_t-\al\vpi_b\right)^2}\>.
\end{equation}

\section{Radiators \label{App:Rad}}
Here we compute the radiators for the total $\Ko$ distribution,
$\cR_\conf(\nu,\be,\gamma)$, to SL accuracy.  The radiator is given by
the combination of dipole contributions,
\begin{equation}
  \label{eq:ARad}
  r_{ab}=\int [dk]\>w_{ab}(k)\sum_{\ell=1}^4\>[1-u_{\ell}]\>\Theta_{\ell}(k)\>,
\end{equation}
with $\Theta_{\ell}(k)$ the support function restricting gluon 
momentum $k$ to the quadrant $C_{\ell}$
and the source functions $u_{\ell}$ defined in \eqref{eq:u}, \eqref{eq:u'}.

Expressing the parton phase space $[dk]$ in terms of Sudakov variables
and invoking the soft distribution $w_{ab}$ in \eqref{eq:dip} we have
\begin{equation}
  \label{eq:Arad}
  r_{ab}\!=\!\int_{-Q}^Q\! dk_x \!\int_{-K_{ym}}^{K_{ym}}\! \frac{dk_y}{\pi} 
 \! \int_0^{\al_m}\!\frac{d\al}{\al} 
\!\left\{
\frac{\as(k_{t,ab})}{\pi}
\frac{\al^2(\vpi_a-\vpi_b)^2}{(\vec{k}_t-\al\vpi_a)^2(\vec{k}_t-\al\vpi_b)^2}
\right\}
\sum_{\ell}[1-u_{\ell}]\>\Theta_{\ell}(k)\>.
\end{equation} 
We recall that $\vpi_a$ are jet transverse direction vectors, 
$\vpi_1=0$ and $\vpi_{2,3}$ given in \eqref{eq:spidef}.

We have three integrations to perform. 
The result of the first two integrals, in $\al$ and $k_y$, has the
structure 
\begin{equation}
  \label{eq:2ints}
\int k_y \int d\al \> \propto\> \frac{\as}{\abs{k_x}}\left(
    \ln\frac{Q^{\,\prime}}{\abs{k_x}} \>+\>\cO{k_x}\right),
\quad \frac{Q^{\,\prime}}{Q}=\cO{1}\,,  
\end{equation}
where we have omitted $\cO{k_x}$ terms as producing non-logarithmic
$\cO{\as}$ corrections upon $k_x$-integration.  (For the same reason
we have set the lower limit of $\al$-integration, $\al\ga (k_t/Q)^2$,
to zero, since the integral converges.)
Similar correction originates from the precise
limit of the $k_x$ integration, $\abs{k_x}\la Q$. Therefore we were
free to set it to $Q$ in \eqref{eq:Arad}.  
\footnote{
We remind the reader that all these $\cO{\as}$ uncertainties, as well
as those we shall encounter below, are related to the coefficient
function~\cite{BSZ}.}

At the same time, the exact limits of $\al$ and $k_y$ integrals do
matter when
they affect the scale of the logarithm, $Q'$, in
\eqref{eq:2ints} which are responsible for the SL correction to the
radiator.
The logarithmic terms in \eqref{eq:2ints} originate from three
``collinear regions'' when the gluon transverse direction vector
$\vpi=\vec{k}_t/\al$ is close to one of $\vpi_a$. 
In the direction $P_1$ it is the $\al$-integration that produces a
collinear logarithm. Noticing that $P_{1z}=TQ/2$, we then set $\al_m$ to
the maximal value of $\al$ kinematically allowed in the
right-hemisphere jet,
\begin{equation}
  \label{eq:alm}
  \al_m\>=\> T\,.
\end{equation}
In the regions collinear to the left-hemisphere jets $a=2,3$,  $\al$
and $k_y$ are linked by the condition $\vpi\simeq\vpi_a$. In this case 
the kinematical limit can be expressed in terms of the $y$-component: 
$P_{2y}=\abs{P_{3y}}=T_MQ/2$ translates into 
\begin{equation}
  \label{eq:Kym}
   K_{ym} \>=\> \frac{T_MQ}{2}\,.
\end{equation}
In \eqref{eq:Arad} we employed soft radiation probabilities $w_{ab}$.
However, in the collinear regions hard parton splitting should be
accounted for which produces another important SL corrections similar
to those coming from the kinematical limits \eqref{eq:alm} and
\eqref{eq:Kym}.  These corrections will be taken care of in the end of
Appendix~\ref{App:d1a}.

We now proceed with successive integrations.  Since the sources $u$ do
not depends on $\al$ we first compute the $\al$-integral.  Then we
compute the $k_y$- and $k_x$-integrals.

\subsection{Integrating over $\al$}
We calculate separately contributions from the right ($R$) and left ($L$)
hemispheres,
\begin{equation}
\label{R/L}
R \>\Rightarrow\>\frac{k_t}{Q} < \al < \al_m= T\>,\qquad L\>\Rightarrow \>
\al < \frac{k_t}{Q} \>,
\end{equation}
and define 
\begin{equation}
\label{I}
\begin{split}
I^{R}_{ab}=\int_{k_t/Q}^{\al_m}\frac{\al d\al\>(\vpi_a-\vpi_b)^2}
{(\vec{k}_t-\al\vpi_a)^2(\vec{k}_t-\al\vpi_b)^2}\>,
\qquad
I^{L}_{ab}=\int_{0}^{k_t/Q}\frac{\al d\al\>(\vpi_a-\vpi_b)^2}
{(\vec{k}_t-\al\vpi_a)^2(\vec{k}_t-\al\vpi_b)^2}\,.
\end{split}
\end{equation}
The $\al$-integration yields
\begin{equation}
\label{intal}
\begin{split}
& \int_0^{\al}\frac{\al d\al\>
(\vpi_a-\vpi_b)^2}{(\vk-\al\vpi_a)^2(\vk-\al\vpi_b)^2}  \\
& =\frac{1}{k_y^2+h_{ab}^2 k_x^2}
\!\left\{\! \frac{h_{ab}}{2}
\ln\!\frac{(\al\spi_a\!-\!k_y)^2+k_x^2}{(\al\spi_b\!-\!k_y)^2+k_x^2}
+\frac{k_y}{|k_x|}\left(\arctan\frac{\al\spi_a\!-\!k_y}{|k_x|}
+\arctan\frac{\al\spi_b\!-\!k_y}{|k_x|}\right)\right\}.
\end{split}
\end{equation}
Here $h_{ab}$ is a function of $\spi_a$ and $\spi_b$ which has the
following geometrical meaning:
\begin{equation}
  \label{eq:hdef}
h_{ab}\>=\> \frac{\spi_a+\spi_b}{\spi_a-\spi_b}
=\frac{\tan\half\theta_a+\tan\half\theta_b}
      {\tan\half\theta_a-\tan\half\theta_b}
= \frac{\sin\half(\theta_a+\theta_b)}{\sin\half(\theta_a-\theta_b)},
\end{equation}
with $\theta_a$ the angle between the momentum $\vec{P}_a$ and the
thrust axis.  We have $h_{a1}=1$ ($a=2,3$) and $h_{23}=\rho$.

In the limit of small aplanarity, $\abs{k_x}\ll Q$, for the dipoles
involving the right-hemisphere jet $P_1$ we derive
\begin{eqnarray}
\label{eq:I1aR}
I^{R}_{1a}&=&
\frac{1}{k_t^2}\left\{ 
\frac12 \ln\frac{\al^2_m\spi_a^2}{\kappa_a^2+k_x^2} 
+\frac{k_y}{\abs{k_x}}\left(\frac{\pi}{2}\left(1-2\vartheta(-\spi_a)\right)
-\arctan\frac{\kappa_a}{\abs{k_x}}
\right)\right\}, \\
\label{eq:I1aL}
I^{L}_{1a}&=&\frac{1}{k_t^2} 
 \left\{\frac12 \ln\frac{\kappa_a^2+k_x^2}{k_t^2} 
 + \frac{k_y}{\abs{k_x}}
\left(\arctan\frac{\kappa_a}{\abs{k_x}}
+ \arctan\frac{k_y}{\abs{k_x}}\right) \right\}.
\end{eqnarray}
For the left-hemisphere dipole we get
\begin{equation}
\label{eq:I23}
\begin{split}
I^{R}_{23}= & 
\frac{1}{k_y^2+ \rho^2k_x^2} 
\cdot \left\{ \frac{\rho}2\ln\frac{\spi_2^2}{\spi_3^2}\>+\> \frac{\rho}2
\ln\frac{\kappa_3^2+k_x^2}{\kappa_2^2+k_x^2} \>-\> \frac{k_y}{\abs{k_x}}
\left(  \arctan\frac{\kappa_2}{\abs{k_x}}
+ \arctan\frac{\kappa_3}{\abs{k_x}}\right) \right\}, \\
I^{L}_{23}=& \frac{1}{k_y^2+ \rho^2k_x^2} 
\cdot \left\{ \frac{\rho}2 \ln\frac{\kappa_2^2+k_x^2}{\kappa_3^2+k_x^2} 
\right.\\& \left. \>+\> \frac{k_y}{\abs{k_x}}
\left(  \arctan\frac{\kappa_2}{\abs{k_x}}
+ \arctan\frac{\kappa_3}{\abs{k_x}} 
+2\arctan\frac{k_y}{\abs{k_x}}\right) \right\}.
\end{split}
\end{equation}
Here
$$
\kappa_a \equiv \frac{k_t}{Q}\spi_a\!-\!k_y\>, \qquad 
\kappa_2>0\>, \quad \kappa_3<0\,.
$$

\subsection{Integrating over $k_y$}
To obtain the $k_x$-integrand we need to perform the $k_y$-integration 
\begin{equation}
\label{eq:rab}
\begin{split}
r_{ab}= \int_{-Q}^{Q} dk_x \frac{\as(2|k_x|)}{\pi}\>B_{ab}(k_x)\>.
\end{split}
\end{equation}
The origin of the running coupling argument in \eqref{eq:rab} is 
explained below in Appendix~\ref{App:running}.

To compute the functions $B_{ab}(k_x)$ we consider two terms
$B_{ab}^{(U)}$ and $B_{ab}^{(D)}$ coming from the upper ($U$) and
lower ($D$) hemispheres, each containing left- and right-hemisphere
contributions,
\begin{equation}
\label{eq:BabU}
B_{ab}^{(U)} 
= \int_0^{K_{ym}}\frac{dk_y}{\pi} \> \left\{
I_{ab}^L\cdot\left[1-u_2\right]\>
+\>I_{ab}^R\cdot\left[1-u_{12}\right]\>\right\},
\end{equation}
and
\begin{equation}
\label{eq:BabD}
B_{ab}^{(D)} 
=\int_{-K_{ym}}^0\frac{dk_y}{\pi} \> \left\{
I_{ab}^L\cdot\left[1-u_3\right]\>
+\>I_{ab}^R\cdot\left[1-u_{13}\right]\>\right\}.
\end{equation}
From \eqref{eq:u} we have
\begin{equation}
  \label{eq:uu}
u_{12}=u(\be_{12},\gamma)\>,\quad
u_{13}=u(\be_{13},-\gamma)\>,\quad
u_{2}=u(\be_{2},0)\>,\quad
u_{3}=u(\be_{3},0)\>,
\end{equation}
where $\be_{1a}=\be_1+\be_a$ and 
\begin{equation}
u(\be,\gamma)=e^{-\nu[\abs{\!k_{x}\!}+i\be k_{x}+i\gamma \abs{\!k_{y}\!}]}\>.
\end{equation}
Only the right-hemisphere sources $u_{1a}$ depend on $k_y$.

\subsubsection{Dipole $23$\label{App:d23}}
Consider first the $23$-dipole contributions $B_{23}^{(U/D)}$.  The DL
piece is obviously contained in the left-hemisphere piece $I^L_{23}$.
At the same time, the contributions involving $I^R_{23}$ are
subleading: their $k_y$-integrals are not enhanced by $\log
\abs{k_x}$.  
Therefore, with SL accuracy the accompanying $[1-u_{1a}]$
factors in \eqref{eq:BabU} and \eqref{eq:BabD} can be simplified, set
equal to $[1-u_a]$ and factored out:
\[
I_{23}^L\cdot[1-u_a]+I_{23}^R\cdot[1-u_{1a}] \>\to\>
 (I_{23}^L+I_{23}^R)\cdot[1-u_{a}]\,.
\]
Then, the terms in $I_{23}^R$ and $I_{23}^L$ due to the boundary between
the $R$- and $L$-hemispheres cancel in the sum, and, using that the
sources are now $k_y$-independent, we obtain
\begin{equation}
\begin{split}
B_{23}^{(U)} &= \left[1-u_2\right]\cdot{\cal L}\,,\qquad
B_{23}^{(D)} \>=\> \left[1-u_3\right] \cdot{\cal L} \,; \\ 
{\cal L}
&= \int_0^{K_{ym}}\frac{dk}{\pi(k^2+ \rho^2k_x^2)}
\left(\frac{\rho}2 \ln\frac{\spi_2^2}{\spi_3^2}\>+ 
\> \frac{2k}{\abs{k_x}} \arctan\frac{k}{\abs{k_x}}\right),
\end{split}
\end{equation}
where $k=k_y$ ($k=-k_y$) in the up (down) hemisphere.  We extract the
logarithmic contribution coming from the $\arctan$ term in the region
$\abs{k_x}\ll k \ll K_m=\half T_MQ$ (see \eqref{eq:Kym}) and using
\begin{equation}
  \label{eq:int}
 \int_0^\infty
 \frac{2k\,dk}{\pi(k^2+\rho^2k_x^2)}\arctan\frac{\abs{k_x}}{k}
 = \ln\frac{1+\rho}{\rho}\,,
\end{equation}
we finally arrive at 
\begin{equation}
  \label{eq:calL}
  {\cal L} = \frac12 \ln\frac{(1-T)Q^2}{4\,k_x^2}\,,
\end{equation}
where we have used the relations 
$$
T_M^2=(1-T)(1-\rho^2)
  \qquad\mbox{and}\quad
 \frac{\spi_2}{|\spi_3|}=\frac{1+\rho}{1-\rho}\,,
$$
following from \eqref{eq:rhodef} and \eqref{eq:spi}, respectively.

\subsubsection{Dipoles $12$ and $13$\label{App:d1a}}
The contributions $1a$ can be simplified in a similar way.  The
left-hemisphere contribution \eqref{eq:I1aL} is accompanied by
$k_y$-independent source $u_a$. The same source can be attributed,
however, to the second term in \eqref{eq:I1aR} as well, since the
$k_y$-integration is non-logarithmic and produces a subleading
contribution.
After this simplification the $R/L$-boundary terms cancel in the sum
of $I^R_{1a}$ and $I^L_{1a}$, and we arrive at
\begin{eqnarray}
\label{eq:B1aU}
 B_{1a}^{(U)} 
\!\!\!\!&=&\!\!\!\! \int_0^{K_{ym}}\frac{dk_y }{\pi k_t^2}
\left( \frac{1}{2}\ln\frac{\al_m^2\spi_a^2}{k_t^2} 
\left[1\!-\!u_{12}\right] 
+ \frac{k_y}{\abs{\!k_x}\!}\left[\frac{\pi}{2}
\sgn(\spi_a) \!+\! \arctan\frac{k_y}{\abs{\!k_x}\!} \right] \!
 \left[1\!-\!u_2\right] \right)\!,  \\ 
\label{eq:B1aD}
 B_{1a}^{(D)} \!\!\!\!&=&\!\!\!\! \int^0_{-K_{ym}}\!\frac{dk_y }{\pi k_t^2}
\left( \frac12\ln\frac{\al_m^2\spi_a^2}{k_t^2} 
\left[1\!-\!u_{13}\right]
+ \frac{k_y}{\abs{\!k_x}\!}\left[ \frac{\pi}{2} \sgn(\spi_a)
\!+\! \arctan\frac{k_y}{\abs{\!k_x}\!} \right] \!
 \left[1\!-\!u_3\right]   \right)\!.
\end{eqnarray}
As before, it is important to keep the dependence of the sources on
the finite parameters $\be$ and $\gamma$ only in the contributions
enhanced by $\ln(Q/\abs{k_x})$. 
Such factor (collinear logarithm due to the jet~\#1) is explicitly
present in the first terms in \eqref{eq:B1aU}, \eqref{eq:B1aD}.
Adding these contributions together and integrating over $k_y$ results
in substituting the R-hemisphere sources by the average source,
$$
\frac{[1-\bar{u}_1]}{2\abs{k_x}}\ln\frac{Q^2}{k_x^2}\>, 
$$
where
\[
1-\bar{u}_1\>\equiv\>\frac1\pi\int_0^\infty \frac{dy}{1+y^2} 
\left[\,(1-u_{12})+(1-u_{13})\,\right], \qquad y\equiv 
\abs{{k_y}/{k_x}}\,.
\]
Collinear enhancement factor due to the jet \#2 originates from
logarithmic $k_y$-integration in the second term of \eqref{eq:B1aU}
(second quadrant), due to the jet \#3 --- in the second term of
\eqref{eq:B1aD} (third quadrant).  Remaining finite pieces can be
evaluated using \eqref{eq:int}.

Setting the limits $\al_m=T$ and $K_{ym}=T_MQ/2$ according to
\eqref{eq:alm} and \eqref{eq:Kym} and using (\ref{eq:spi}), we finally
obtain
\begin{equation}
\label{eq:BB}
B_{1a}= B_{1a}^{(U+D)} = 
\frac{(1-\bar{u}_1)}{2\abs{k_x}} \ln\frac{Q^2}{k_x^2} 
+\frac{(1-u_a)}{2\abs{k_x}} \ln\frac{Q^2}{k_x^2}
 \>+\> \frac{(1-u_0)}{\abs{k_x}} \ln\frac{T T_M|\spi_a|}{8 Q}\>,
\end{equation}
where $u_0$ in the last subleading term is a source whose $\be$-,
$\gamma$-dependence can be chosen arbitrary.  Using this freedom we
can absorb the last term in \eqref{eq:BB} into rescaling of the first
two namely,
\begin{equation}
\label{eq:B1ahs}
\ln\frac{Q^2}{k_x^2} \>\to\> \ln\frac{T T_M|\spi_a|\,Q}{8\,k_x^2}\,.
\end{equation}
We observe that the hard scales in \eqref{eq:calL} and
\eqref{eq:B1ahs} have a simple geometrical interpretation. Indeed,
\begin{equation}
  \label{eq:scales}
(1-T)Q^2\>=\>2P_2P_3=Q^2_{23}\>,
\qquad
\frac{T T_M|\spi_a|Q}{2}=2P_1P_a=Q^2_{1a}\>,
\end{equation}
with $Q_{ab}$ the invariant dipole masses, see \eqref{eq:Qa}.

Our analysis was based up to now on the soft radiation matrix element,
\eqref{eq:dip}.  To fully take into account SL effects from the region
of large secondary parton momenta, we have, in addition to fixing the
upper limits $\al_m$ and $k_{ym}$, to consider also hard collinear
parton splitting.  Due to collinear factorization, these corrections
are process-independent and can be easily taken into account by proper
rescaling of jet hardness parameters.  They amount to supplying the
invariant dipole masses by the factors
\begin{equation}
  \label{eq:gdef}
 Q^2_{ab} \>\to\> Q^2_{ab} \cdot 
\exp\left\{ - \half(g_a+g_b) \right\}, \qquad
g_a=\left\{
\begin{array}{cl}
\frac{3}{2} & \mbox{for a quark/antiquark}\,, \\[2mm]
\frac{\be_0}{2N_c}  & \mbox{for a gluon}\,.
\end{array}  \right. 
\end{equation}
We finally obtain
\begin{equation}
\begin{split}
B_{23} &=\frac{\left[(1-u_2)+(1-u_3)\right]}{2\,\abs{k_x}}
\ln\frac{Q_{23}^2e^{-\half(g_2+g_3)}}{4k_x^2}\,, \\
B_{1a} &=\frac{\left[(1-\bar{u}_1)+(1-u_a) \right]}{2\,\abs{k_x}}
\ln\frac{Q_{1a}^2e^{-\half(g_1+g_a)}}{4k_x^2}\,.
\end{split}
\end{equation}

\subsection{Radiator by assembling bits and pieces}
Now that the $\al$- and $k_y$-integrations have been performed, we are
in a position to assemble the full radiators for three jet
configurations:
\begin{eqnarray}
\label{eq:ARadT3}
 \cR_3&=&\int_{-Q}^{Q} dk_x\frac{\as(2|k_x|)}{\pi}
\frac{N_c}{2}\left(B_{13}+B_{23}-\frac{1}{N^2_c}B_{12}\right), \\
\label{eq:ARadT2}
 \cR_2&=&\int_{-Q}^{Q} dk_x\frac{\as(2|k_x|)}{\pi}
\frac{N_c}{2}\left(B_{12}+B_{23}-\frac{1}{N^2_c}B_{13}\right), \\
\label{eq:ARadT1}
 \cR_1&=&\int_{-Q}^{Q} dk_x\frac{\as(2|k_x|)}{\pi}
\frac{N_c}{2}\left(B_{12}+B_{13}-\frac{1}{N^2_c}B_{23}\right).
\end{eqnarray}
The answers assume a simple when expressed in terms of the invariant
transverse momentum of the hard parton $P_a$ with respect to the
$bc$-dipole
\[ 
p^2_{ta}\equiv \frac{Q^2_{ba}Q^2_{ac}}{Q^2_{bc}}\>.
\]
It is straightforward to verify that, to SL accuracy, the answer can
be represented in a symmetric form as
\begin{equation}
  \label{eq:AradT}
  \cR_\conf\>\!=\!\!\int_{-Q}^Q\!\frac{dk_x}{k_x}\frac{\as(2|k_x|)}{2\pi}\>
\!\Big(
C^{(\conf)}_1\ln\!\frac{Q_1^2}{k^2_x}[1-\bar{u}_1]+
C^{(\conf)}_2\ln\!\frac{Q_2^2}{k^2_x}[1-u_2]+
C^{(\conf)}_3\ln\!\frac{Q_3^2}{k^2_x}[1-u_3]\!\Big),
\end{equation}
where the hard scales are given by:
\begin{equation}
\begin{split}
   Q_{1}^2 =  \frac{p_{t1}^2}{4}e^{-g_1}\>, \qquad
   Q_{2}^2 =  \frac{p_{t2}^2}{4}e^{-g_2}\>, \qquad
   Q_{3}^2 =  \frac{p_{t3}^2}{4}e^{-g_3}\>.
\end{split}
\end{equation}
The rule for the colour factors in \eqref{eq:AradT} is simple:

\begin{equation}
\label{eq:crule}
 C^{(a)}_a=N_c\,; \qquad   C^{(a)}_b=C_F\,, \quad \mbox{for}\> a\ne b\,,
\end{equation}

and the hard-splitting rescaling factors $e^{-g_a}$ are defined in
\eqref{eq:gdef}.

The universal representations \eqref{eq:ARadT3}--\eqref{eq:ARadT1} do
not have a smooth 2-jet limit.  In subsection~\ref{sec:2jet}
alternative formulae are presented which are equivalent to the
previous ones in the 3-jet kinematics, $T_M\la T=\cO{1}$, but are
better behaved when the system assumes a quasi-two-jet kinematics,
$T_M\ll T$.

\subsection{Radiator for the right distribution \label{App:RadR}}
In this case the source does not depend on $k_y$, and we have
\begin{equation}
  \label{eq:ARdaR}
\begin{split}
 & \cR^R_{ab}(\nu,\be)=\int_{-Q}^{Q}dk_x\>
 \frac{\as(2|k_x|)}{\pi}\>B_{ab}^R(k_x)\>, 
\\& B_{ab}^R(k_x)= \left[1-e^{-\nu(|k_x|+i\be k_x)}\right]\cdot
\int_{-\infty}^{\infty}\frac{dk_y}{\pi}\> I^R_{ab}\>,
\end{split}   
\end{equation}
where the functions $I^R_{ab}$ are given in \eqref{eq:I1aR} and
\eqref{eq:I23}.  
The DL contribution is contained in the first term of $I^R_{1a}$ in
\eqref{eq:I1aR}.  Extracting the large logarithm $\ln(Q^2/k_x)$ which
is embodied into the DL function $r(\bmu,Q^2)$ in \eqref{eq:R1a}, we
calculate the geometry-dependent SL correction factors denoted there
by $F_{ab}$.

The $k_y$ integrals are convergent so we have set the upper limit
$K_{ym}\to\infty$.  Introducing the ratio of momenta $t=k_y/\abs{k_x}$
we obtain the following expressions:
\begin{equation}
  \label{eq:Fabr}
\begin{split}
F_{1a}(\spi_a) &= 2\ln\frac{\al_m\abs{\spi_a}}{Q}\>+\>
 2\>\int_{-\infty}^{\infty}
 \frac{dt}{\pi(1+t^2)}\left\{-\frac{1}{2}\>
 \ln\left(1+\kappa_a^{'2}\right) \right.\\ 
 &\left. +\>t\left(\frac{\pi}{2}\sgn(\spi_a)
 -\arctan(\kappa_a')\right)\right\} , \\
F_{23}(\spi_2,\spi_3) &=2\ln\frac{\spi_2}{\abs{\spi_3}}\>+\>
 2\>\int_{-\infty}^{\infty}\frac{dt}{\pi(\rho^2+t^2)}\left\{\frac{\rho}{2}
 \ln\left(\frac{1+\kappa_3^{'2}}{1+\kappa_2^{'2}}\right)
 \right.\\
 &-\left. t\left(\arctan(\kappa_2')+ 
 \arctan(\kappa_3')\right) \right\}.
\end{split}   
\end{equation}
where
$\kappa_a'\>\equiv\>\frac{\kappa_a}{|k_x|}=\sqrt{1+t^2}\frac{\spi_a}{Q}-t$.
We remind the reader that convergence of the integrals is assured by
$\spi_2>1$, $\spi_3<-1$.

\subsection{Running coupling \label{App:running}} 
The two-loop analysis \cite{BDMZ} entails that the argument of the
running coupling for the dipole distribution $r_{ab}$ in
\eqref{eq:ARad} is given by $k_{t,ab}$, the invariant gluon transverse
momentum with respect to the $ab$-dipole (see\eqref{eq:wij}).
We show here that, to SL accuracy, we can the scale at the value
$2|k_x|$ instead, as has been stated in \eqref{eq:rab}.  This
effective value is obtained after integrating $\as(k_{t,ab})$ over
$\al$ and $k_y$.
To show this we first observe that we need to control the precise
argument of $\as$ only in the DL contributions which originate from
the phase space regions where the gluon is collinear to one of the
three hard partons $P_a$.

Consider first the case of the contribution of $r_{1a}$ from the right
hemisphere.  Here the soft gluon is close to the thrust axis, $P_1$,
so that the invariant transverse momentum reduces to the usual
2-dimensional momentum, $k_{t,1a}^2\simeq k_t^2=k_x^2+k_y^2$, which is
$\al$-independent.  This allows us to perform the $\al$-integration
and obtain $I^R_{1a}$ in \eqref{eq:I1aR}.

Now, to determine the effective scale of $\as$ it suffices to consider
$k_y$-integral of the leading piece of $I_{1a}^R$ proportional to
$\ln(Q^2/k_x^2)$ and integrate over $k_y$.  We have an integral of the
type
\begin{equation}
A= \int_{-K_{ym}}^{K_{ym}}\frac{dk_y}{\pi k_t^2}\>\as(k_t)
  =\frac{1}{k_x}\int_{-y_m}^{y_m}\frac{dy}{\pi(1+y^2)}\>
   \as(k_x\sqrt{1+y^2})\>,
\end{equation}
with $k_x$ positive, $y\equiv k_y/k_x$.  For small $k_x$ the upper
limit of the $y$-integral is large, $y_m=K_{ym}/k_x=T_MQ/2k_x\gg1$,
and can set infinite since the integral converges.  
Expanding the coupling to the first order, we get
\begin{equation}
\begin{split}
k_x\cdot A\>&=\>   \int_{-\infty}^{\infty}\frac{dy}{\pi(1+y^2)}\>
  \left(\as(k_x)-\frac{\be_0}{4\pi}\as^2(k_x)\ln(1+y^2)\right)\>
   +\>\cO{\as^3}\\ 
&\simeq\>
  \left(\as(k_x)-\frac{\be_0}{4\pi}\as^2(k_x)\ln 4\right)
  \>\simeq\> \as(2k_x) \>.
\end{split}
\end{equation}
The regions collinear to $P_2$ or $P_3$ seems more complicated since
here $k_{t,ab}$ depend both on $\al$ and $k_y$.  However, a similar
analysis can be carried out in terms of the Sudakov variables with
$P,\bar{P}$ aligned with the emitting parton, we obtain the same
result for the argument of $\as$.

\subsection{The sources and the $k_x$ integrals \label{App:theta}}
Here we prove the substitution rule \eqref{theta}.  The general
structure of the radiators is
\begin{equation}
 D= \int_0^{{Q}} \frac{dk_x}{k_x} \ln\frac{Q'}{k_x} [\,1-u\,]\,,
\qquad
u \equiv e^{-\nu k_x}\cos(\nu\beta k_x) e^{i\>\nu \gamma y k_x}\,,
\end{equation}
with $Q'=\cO{Q}$ a hard scale.  This we write as
\begin{equation}
\label{eq:D}
 D=  \int_{1/\bmu}^Q \frac{dk_x}{k_x} \ln\frac{Q'}{k_x}
   + \Delta\,, \quad  
\Delta  =\left. \frac{-\partial}{\partial \eps}\left\{
\int_0^{1/\bmu} \frac{dk_x}{k_x} \left(\frac{k_x}{Q'}\right)^{\eps}
-\int_0^{{Q}} \frac{dk_x}{k_x} \left(\frac{k_x}{Q'}\right)^{\eps}\cdot u
\right\}\right|_{\eps=0},
\end{equation}
and optimise the choice of $\bmu$ such that $\Delta=\cO{1}$, i.e.\ it
does not contain a $\ln \nu$-enhancement.  Since $\nu Q\gg1$, the
second integral containing the exponential source function,
$u\propto\exp(-\nu k_x)$, can be safely extended to infinity.

Evaluating the expression in the curly brackets up to $\cO{\eps}$, we
obtain
\begin{equation}
\begin{split}
 \Big\{\>\Big\} &= \eps^{-1}\left[
\left({\bmu}{Q'}\right)^{-\eps} - \Gamma(1+\eps)
\left({\nu}{Q'}\right)^{-\eps}\cdot\half 
\left((1-i\gamma y-i\be)^{-\eps} +(1-i\gamma y+i\be)^{-\eps}\right)
\right] \\
&= - \left({\nu}{Q'}\right)^{-\eps} 
\left( \ln\frac{\bmu}{\nu} -\gamma_E 
-\half\ln\left[(1-i\gamma y)^2+\be^2\right] \>+\> \cO{\eps}\right). 
\end{split}
\end{equation}
Taking the $\eps$-derivative in \eqref{eq:D} we obtain a large
parameter $\ln(\nu Q')$ accompanied by the factor which we set equal
to zero to optimize the choice of $\bmu$:
\begin{equation}
  \bmu = \nu\, e^{\gamma_E} \sqrt{(1-i\gamma y)^2+\beta^2}\>.
\end{equation}
This means that, within SL accuracy, the source factor $[1-u]$ can be
substituted by
\begin{equation}
  [1-u] \to \vartheta(k_x-\bmu^{-1})\,. 
\end{equation}
Performing this substitution in \eqref{eq:AradT} we get the result
reported in the text, see \eqref{eq:Radfin}.

\section{Evaluation of $\cF$ \label{App:cF}}
The expression for $\cF_{\conf}$ is rather complicated.  
Invoking \eqref{eq:I} we split the $\gamma$-integral into two pieces,
\begin{equation}
  \cF \>=\> \cF_r + \cF_i\>, 
\end{equation}
namely the principal value and the $\delta(\gamma)$ contributions,
\begin{equation}
  \frac{1}{\gamma\mp i\eps}= \frac{\mbox{P}}{\gamma} 
      \>\pm\> i\pi\delta(\gamma)\>.
\end{equation}
We have
\begin{equation}
\begin{split}
\cF=\int_{-\infty}^\infty \frac{d\be_2}
                  {\pi (1+\be_2^2)^{1+\half C^{(\conf)}_2  r'}} 
    \int_{-\infty}^\infty \frac{d\be_3}{\pi(1+\be_3^2)^{1+\half
        C^{(\conf)}_3 r'}}
    \int_{-\infty}^\infty \frac{d\be_1}{\pi} \cdot (\cI_r+\cI_i)\,,
\end{split}
\end{equation}
where the two integrands are 
\begin{equation}
\begin{split}
& 
\cI_i\!= \frac{1}{2} \> \left(\frac1{1+\be_{12}^2}\! +\!\frac1{1+\be_{13}^2}\right) 
\left(\sqrt{1+\be_{12}^2}
\sqrt{1+\be_{13}^2}\right)^{\frac{C^{(\conf)}_1 r'}{2}}\>,\\
&
\cI_r\!=\! \left(\frac1{1+\be_{12}^2}\! -\!\frac1{1+\be_{13}^2}\right) 
\!\!\int_0^\infty \!\!\frac{d\gamma}{\pi} 
\!\left(\!\sqrt{(1\!+\!\gamma)^2\!+\!\be_{12}^2}
\sqrt{(1\!+\!\gamma)^2\!+\!\be_{13}^2}\!\right)^{\frac{-C^{(\conf)}_1r'}{2}} 
\!\!\frac{\sin(C^{(\conf)}_1\,r'A_1)}{\gamma}\,.
\end{split}
\end{equation}
Here
\begin{equation}
  A_1 \>=\> \frac1{2\pi} \int_0^\infty \frac{dx}{1-x^2}
  \left[\, \ln\frac{(1+\gamma x)^2+\be_{12}^2}{(1+\gamma)^2+\be_{12}^2}  
 -\ln\frac{(1+\gamma x)^2+\be_{13}^2}{(1+\gamma)^2+\be_{13}^2}\,\right]
\end{equation}
Where as before the colour factors are given by \eqref{eq:crule}.

\subsection{$\cF$ in the first order \label{App:cF1}}
In the $\cO{r'}$ approximation we have, for $\cF_r$
\begin{equation}
    \begin{split}
 \cF_r(\bnu) \simeq C^{(\conf)}_1\, r'\!
 \int_{-\infty}^\infty \frac{d\be_2}{\pi (1\!+\!\be_2^2)} 
   \! \int_{-\infty}^\infty \frac{d\be_3}{\pi(1\!+\!\be_3^2)}
  \!\int_{-\infty}^\infty \frac{d\be_1}{\pi}
\! \left(\frac1{1\!+\!\be_{12}^2}\!-\!\frac1{1\!+\!\be_{13}^2}\right)\>
 \!\frac{1}{\pi} \!\int_0^\infty\! \frac{d\gamma}{\gamma} A_1\>.
\end{split}
\end{equation}
Using
\begin{equation}
    \begin{split}
\frac{1}{\pi}\int_{0}^{\infty} \frac{d\gamma}{\gamma}\> A_1
=\frac{1}{8}\ln\frac{1+\beta_{12}^2}{1+\beta_{13}^2}\>,
 \end{split}
\end{equation}
we find
\begin{equation}
    \begin{split}
 \cF_r(\bnu) &\!\simeq\! \frac{r'C^{(\conf)}_1}{2}\! 
 \int_{-\infty}^\infty \!\frac{d\be_2}{\pi (1\!+\!\be_2^2)} 
    \int_{-\infty}^\infty \frac{d\be_3}{\pi(1\!+\!\be_3^2)}
  \int_{-\infty}^\infty \frac{d\be_1}{2\pi}
 \left(\frac1{1\!+\!\be_{12}^2}\! -\!\frac1{1\!+\!\be_{13}^2}\right) 
\frac12\ln\frac{1\!+\!\beta_{12}^2}{1\!+\!\beta_{13}^2}\>.
 \end{split}
\end{equation}
Using
\begin{equation}
\label{ln4:1}
\int_{-\infty}^\infty \frac{d\be}{\pi}  \frac{\ln(1+\be^2)}{1+\be^2}
= \ln 4\>.
\end{equation}
and 
\begin{equation}
\label{ln4:2}
  \begin{split}
   & \int_{-\infty}^\infty \frac{d\be_{13}}{\pi} \ln(1\!+\!\be_{13}^2)
 \int_{-\infty}^\infty \frac{d\be_{12}}{\pi (1\!+\!\be_{12}^2)} 
 \int_{-\infty}^\infty \frac{d\be_1}{2\pi}
 \frac{1}{1+(\be_{12}\!-\!\be_1)^2} \frac{1}{1+(\be_{13}\!-\!\be_1)^2}
 \> =\>\ln 4\>.
  \end{split}
\end{equation}
we obtain
\begin{equation}
 \cF_r(\bnu) \simeq -\frac{1}{2}\ln4\> 
\frac{C^{(\conf)}_1}{2}\cdot  r'(\bnu)\>.
\end{equation}
Expanding $\cF_i$ in $r'$ we obtain in the first order:
\begin{equation}
\begin{split}
 \cF_i(\bnu) &= 1-\half r'\cdot 
 \int_{-\infty}^\infty \frac{d\be_2}{\pi (1+\be_2^2)} 
    \int_{-\infty}^\infty \frac{d\be_3}{\pi(1+\be_3^2)}
 \int_{-\infty}^\infty \frac{d\be_1}{2\pi}
\left(\frac1{1+\be_{12}^2} +\frac1{1+\be_{13}^2}\right) \\
& \left( C^{(\conf)}_2\ln(1+\be_2^2) + C^{(\conf)}_3\ln(1+\be_3^2) 
+\half  C^{(\conf)}_1
\left[\, \ln(1+\be_{12}^2)+\ln(1+\be_{13}^2)\,\right] \right)
\end{split}
\end{equation}
Using \eqref{ln4:1} and \eqref{ln4:2} we get:
\begin{equation}
\label{cfone}
\cF_i(\bnu) 
= 1-\half\ln 4\left(\frac32 C^{(\conf)}_1+C^{(\conf)}_2+
 C^{(\conf)}_1\right)\cdot r'(\bnu)\>.
\end{equation}
So, in the first order,
\begin{equation}
\label{cf1}
\cF=   \cF_r(\bnu)+\cF_i(\bnu) 
= 1-\half \ln 4 \left(2\,C^{(\conf)}_1+C^{(\conf)}_2+
C^{(\conf)}_3\right)\cdot r'(\bnu)
\end{equation}


\begin{thebibliography}{99}
\bibitem{thrust} 
%RESUMMATION OF LARGE LOGARITHMS IN E+ E- EVENT SHAPE DISTRIBUTIONS. 
  S.~Catani, L.~Trentadue, G.~Turnock and B.R.~Webber,
%  \Journal{\NPB}{407}{3}{1993}. 
\npb{407}{1993}{3}.

\bibitem{Cpar} 
%RESUMMED C PARAMETER DISTRIBUTION IN E+ E- ANNIHILATION.
 S. Catani and B.R. Webber, 
%\Journal{\PLB}{427}{377}{1998} [hep-ph/9801350]. 
\plb{427}{1998}{377} [hep-ph/9801350]. 

\bibitem{broad} 
% JET BROADENING MEASURES IN E+ E- ANNIHILATION.
S.~Catani, G.~Turnock and B.R.~Webber, 
%\Journal{\PLB}{295}{269} {1992},\\
\plb{295}{1992}{269};\\
   Yu.L. Dokshitzer, A. Lucenti, G. Marchesini and G.P. Salam, 
%   \Journal{\JHEP}{01}{011}{1998} [hep-ph/9801324]. 
\jhep{01}{1998}{011}\\%  
{}[hep-ph/9801324]. 

\bibitem{thrustDIS} 
   V.~Antonelli, M.~Dasgupta and G.P.~Salam, 
%   \Journal{\JHEP}{0002}{001}{2000}  [hep-ph/9912488]. 
\jhep{02}{2000}{001} [hep-ph/9912488]. 

\bibitem{1/Q} M.~Beneke, 
% RENORMALONS.
% \Journal{\PREP}{317}{1} {1999}  [hep-ph/9807443],  \\              
\prep{317}{1999}{1}  [hep-ph/9807443];  \\      
B.R.~Webber,  
%\Journal{\NPPS}{71}{66} {1999}  [hep-ph/9712236] 
\npps{71}{1999}{66} [hep-ph/9712236]. 
%RENORMALON PHENOMENA IN JETS AND HARD PROCESSES.
% Talk given at 27th International Symposium 
% on Multiparticle Dynamics (ISMD 97), Frascati, Italy, 8-12 Sep
% 1997. 

\bibitem{Blois}
Yu.L. Dokshitzer, {\em Perturbative QCD and Power Corrections},
Invited talk at 11th Rencontres de Blois: Frontiers of Matter, Chateau
de Blois, France, 28 Jun -- 3 Jul 1999 [hep-ph/9911299].

\bibitem{AO} %%% AO
B.I.~Ermolayev and V.S.~Fadin, 
%\Journal{\JETPL}{33}{285}{1981},\\
\jetpl{33}{1981}{269};\\
A.H.~Mueller, 
%\Journal\PLB{104}{161}{1981}.
\plb{104}{1981}{161}.

\bibitem{SAO} %strict AO 
%
  Yu.L.~Dokshitzer and S.I.~Troyan, ``{\em Asymptotic freedom and Local
  Parton-Hadron Duality in QCD jet physics}'',
  Proceedings of 19th Leningrad Winter School, Leningrad 1984, p.~144;
%  (in Russian); 
for review see 
  Yu.L.~Dokshitzer, V.A.~Khoze, A.H.~Mueller and S.I.~Troyan,
  ``Basics of Perturbative QCD'',  
  ed. J. Tran Thanh Van, Editions Fronti\`eres, Gif-sur-Yvette, 1991; \\
% 
  A.H.~Mueller, 
%\Journal{\NPB}{213}{85}{1983}, Erratum quoted
\npb{213}{1983}{85},  Erratum quoted
  \ibid{B~241}{1984}{141}; \\
  E.D.~Malaza and B.R.~Webber,  
%\Journal\PLB{149}{501}{1984};
\plb{149}{1984}{501}.

\bibitem{mm3jet}
%COLLECTIVE QCD EFFECTS IN THE STRUCTURE OF FINAL MULTI - HADRON STATES. 
Yu.L.~Dokshitzer, S.I.~Troyan and  V.A.~Khoze, 
%\Journal{\SJNP}{46}{712}%-719,
%{1987}.    
\sjnp{46}{1987}{712}.
%%%%% Yad.Fiz.46:1220-1232,1987 
% %
% %MULTIPLE HADROPRODUCTION IN HARD PROCESSES 
% %WITH NONTRIVIAL TOPOLOGY. (IN RUSSIAN).
% Yu.L.~Dokshitzer, S.I.~Troyan and V.A.~Khoze 
% \Journal{\SJNP}{47}{881}
% %-888,
% {1988},\\ %%%%% Yad.Fiz.47:1384-1396,1988 

\bibitem{BSZ} 
%  G.P.~Salam and G.~Zanderighi, BICOCCA-FT-99-28, Jul
%  1999.  \\
%  To be published in the proceedings of High Energy Physics
%  International Euroconference on Quantum Chromo Dynamics - QCD '99,
%  Montpellier, France, 7-13 Jul 1999 [hep-ph/9909324], \\
  A.~Banfi, G.P.~Salam and G.~Zanderighi, work in progress. 


\bibitem{l3col}
L3 Collaboration, B.~Adeva et al., 
% \Journal{\ZPC}{55} {39}{1992},\\
\zpc{55}{1992}{39}; \\
D.P~.Barber, et al., 
%\Journal{\PLB}{89}{139}{1979}.
\plb{89}{1979}{139}.


\bibitem{broadNP} 
%REVISITING NONPERTURBATIVE EFFECTS IN THE JET BROADENINGS.
Yu.L. Dokshitzer, G.~Marchesini and G.P.~Salam, 
%\Journal{\epj}{3}{1} {1999}  [hep-ph/9812487].
\epj{3}{1999}{1}  \\%
{}[hep-ph/9812487].

\bibitem{KoNP} A.~Banfi, Yu.L. Dokshitzer, G.~Marchesini and
  G.~Zanderighi, under preparation.  %(non-PT $\Ko$)
\bibitem{Catani} S.~Catani and  M.~Grazzini, [hep-ph/9908523] 
% INFRARED FACTORIZATION OF TREE LEVEL QCD AMPLITUDES 
% AT THE NEXT-TO-NEXT-TO-LEADING ORDER AND BEYOND.

\bibitem{BDMZ} 
 A.~Banfi, Yu.L. Dokshitzer, G.~Marchesini and
  G.~Zanderighi, under preparation. %\\ (2-loop future)

\bibitem{DLMS} 
% UNIVERSALITY OF 1 / Q CORRECTIONS TO JET SHAPE OBSERVABLES RESCUED. 
Yu.L.~Dokshitzer, A.~Lucenti, G.~Marchesini and G.P.~Salam, 
%\Journal{\NPB}{511}{396}%-418,
%{1998} [hep-ph/9707532].
\npb{511}{1998}{396}  [hep-ph/9707532].

\bibitem{CMW} 
% QCD COHERENT BRANCHING AND SEMIINCLUSIVE PROCESSES AT LARGE X.
S.~Catani,  G.~Marchesini and B.R.~Webber,
%\Journal\NPB{349}{635}%-654,
%{1991}. 
\npb{349}{1991}{635}. 

\bibitem{Milan} 
 % ON THE UNIVERSALITY OF THE MILAN FACTOR 
 % FOR 1 / Q POWER CORRECTIONS TO JET SHAPES.
Yu.L. Dokshitzer, A.~Lucenti, G.~Marchesini and G.P.~Salam, 
%\Journal{\JHEP}{05}{003}{1998} [hep-ph/9802381], \\  
\jhep{05}{1998}{003} \\%
{}[hep-ph/9802381]; \\  
% TWO LOOP ENHANCEMENT FACTOR FOR 1 / Q CORRECTIONS 
% TO EVENT SHAPES IN DEEP INELASTIC SCATTERING.
M.~Dasgupta and B.R.~Webber,
%\Journal{\JHEP}{10}{001}{1998} [hep-ph/9809247].
\jhep{10}{1998}{001} [hep-ph/9809247].

\bibitem{KO}  
%PERTURBATIVE QCD APPROACH TO MULTIPARTICLE PRODUCTION.
  V.A.~Khoze and W.~Ochs, 
%  \Journal{\IJMPA}{12}{2949}%-3120,
%{1997}  [hep-ph/9701421].
\ijmpa{12}{1997}{2949}  [hep-ph/9701421].
\end{thebibliography}
\end{document}